\documentclass[prl,groupedaddress,reprint,superscriptaddress,floats,nofootinbib,preprintnumbers]{revtex4-1}
\usepackage{graphicx} 
\usepackage{float}
\usepackage{amsmath}
\usepackage{epstopdf}
\usepackage{xspace}
\usepackage{rotating}
\usepackage{longtable}
\usepackage{multirow}
\usepackage{booktabs}
\usepackage[breaklinks=true]{hyperref}
\hypersetup{
  colorlinks=true,
  linkcolor=blue,
  citecolor=blue,
  urlcolor=blue
}

\graphicspath{{./figs/}}
\usepackage{array,tabularx,epsfig,mathrsfs,rotating}
\usepackage{ifthen}
\usepackage{amsfonts}
\usepackage{amssymb}
\usepackage{slashed}

\usepackage[dvipsnames]{xcolor}

\newcommand{\met}{\mbox{$\protect \raisebox{.3ex}{$\not$}E_T$}}

\begin{document}

\begin{flushleft} 
IFT-UAM-CSIC-22-55  
\end{flushleft}
\begin{flushright} 
FTUAM-22-1 
\end{flushright}

\title{\boldmath
Semi-dark Higgs decays: sweeping the Higgs neutrino floor 
}

\author{J.~A. Aguilar-Saavedra}
\email[]{ja.a.s@csic.es}
\affiliation{Instituto de Fisica Teorica, IFT-UAM/CSIC,
Cantoblanco, 28049, Madrid, Spain}

\author{J.~M. Cano}
\email[]{josem.cano@uam.es}
\affiliation{Instituto de Fisica Teorica, IFT-UAM/CSIC,
Cantoblanco, 28049, Madrid, Spain}
\affiliation{Departamento de Fisica Teorica, Universidad Autonoma de Madrid,
Cantoblanco, 28049, Madrid, Spain
}

\author{D.~G. Cerdeño}
\email[]{davidg.cerdeno@gmail.com}
\affiliation{Instituto de Fisica Teorica, IFT-UAM/CSIC,
Cantoblanco, 28049, Madrid, Spain}
\affiliation{Departamento de Fisica Teorica, Universidad Autonoma de Madrid,
Cantoblanco, 28049, Madrid, Spain
}

\author{J.~M. No}
\email[]{josemiguel.no@uam.es}
\affiliation{Instituto de Fisica Teorica, IFT-UAM/CSIC,
Cantoblanco, 28049, Madrid, Spain}
\affiliation{Departamento de Fisica Teorica, Universidad Autonoma de Madrid,
Cantoblanco, 28049, Madrid, Spain
}

\hfill\draft{ }

\begin{abstract}
We study exotic Higgs decays $h \to Z X$, with $X$ an invisible beyond the Standard Model (SM) particle, resulting in a semi-dark final state. Such exotic Higgs decays may occur in theories of axion-like-particles (ALPs), dark photons or pseudoscalar mediators between the SM and dark matter. The SM process $h\to Z\nu\bar{\nu}$ represents an irreducible ``neutrino floor'' background to these new physics searches, providing also a target experimental sensitivity for them. We analyze $h \to Z + \textsl{invisible}$ searches at the LHC and a future ILC, showing that these exotic Higgs decays can yield sensitivity to unexplored regions of parameter space for ALPs and dark matter models. 
\end{abstract}

\maketitle

\noindent {\bf Introduction.}~The Higgs boson discovered at the Large Hadron Collider (LHC) offers a unique window into new physics, and it is paramount to study its properties with precision. Exotic Higgs decays, 
i.e.~decays of the Higgs boson not present in the Standard Model (SM), constitute a primary avenue to probe the existence of new physics~\cite{Curtin:2013fra}. In the last years, there has been an intense experimental program at the LHC to search for such exotic Higgs decays~\cite{CMS:2021pcy,ATLAS:2021ldb,CMS:2020ffa,CMS:2018qvj,CMS:2019spf,ATLAS:2020ahi,ATLAS:2021hbr,CMS:2018nsh,CMS:2018zvv,ATLAS:2018pvw,ATLAS:2018jnf,ATLAS:2020pcy} (see also~\cite{Cepeda:2021rql} and references therein). Such searches have mainly targeted either fully visible final states, e.g.~$h\to 2 f \,2 f' $ (with $f, f'$ SM fermions) or a fully invisible Higgs decay (so-called \textit{invisible Higgs width}). 
Considering all/part of the Higgs boson decay products in these exotic decays to be invisible at colliders is well-motivated theoretically, e.g.~if the Higgs boson directly interacts with a dark (i.e.~not feeling the SM gauge interactions) sector of Nature, possibly containing the dark matter (DM) particle(s), or if the Higgs decay products are very long-lived and decay outside the LHC detectors. Yet, partially invisible (\textit{semi-dark}) Higgs boson decays constitute a much less explored avenue to search for new physics beyond the SM (BSM) coupled to the Higgs boson, and studies of these semi-dark Higgs decays exist in the literature for very few BSM scenarios~\cite{Englert:2012wf,Petersson:2012dp} (see~\cite{CMS:2015ifd,CMS:2019ajt,CMS:2020krr,ATLAS:2021pdg,ATLAS:2021edm} for existing experimental searches). 
Such searches are fully complementary to searches for invisible Higgs decays, and generally probe different regions of parameter space of the same BSM theories. Semi-dark Higgs decays allow in particular to obtain key information on the nature of the coupling between the Higgs and the invisible state(s), by reconstructing the visible part of the exotic Higgs decay. 

In this work we target the previously unexplored semi-dark Higgs decay $h \to Z X$, with $X$ a BSM particle invisible at the LHC (manifesting as missing transverse energy $\slashed{E}_T$), and we show it is a promising avenue to probe various well-motivated BSM scenarios:~$X$ could be an axion like particle (ALP) or dark photon that decays invisibly or is long-lived, escaping the detector. It could also be a pseudoscalar mediator particle between the SM and a dark sector of Nature containing the DM particle\footnote{A pseudoscalar mediator would nicely explain the absence of a spin-independent signal in current DM direct detection experiments~\cite{XENON:2018voc}. These \textit{pseudoscalar portal to DM} scenarios have also been proposed to explain~\cite{Boehm:2014hva,Izaguirre:2014vva,Ipek:2014gua} the 
$\gamma$-ray excess~\cite{Goodenough:2009gk,Hooper:2010mq} in the Fermi-LAT observations of the Milky Way Galactic Center~\cite{TheFermi-LAT:2015kwa}.}.

We focus our study on the leptonic decay of the $Z$ boson, $Z\to \ell\ell$ (with $\ell = e, \mu$), leading to the Higgs final state $h\to Z X \to \ell\ell + \slashed{E}_T$.
Incidentally, the SM decays $h \to Z Z^* \to \ell\ell \nu\bar{\nu}$ and $h \to W W^* \to \ell \nu \ell \bar{\nu}$ yield the same final state. For the latter, the two leptons do not reconstruct the $Z$ boson mass $m_Z \simeq 91$ GeV in general, which can be used to tell apart $h \to Z X$ from this SM decay process. However, $h \to Z Z^* \to \ell\ell \nu\bar{\nu}$ completely mimics a possible BSM signal. The SM decay $h \to Z \nu\bar{\nu} $ 
then constitutes a ``\textit{neutrino floor}"\footnote{In analogy to DM direct detection experiments, where coherent elastic neutrino-nucleus scattering can pose an irreducible background to DM searches, known as the ``neutrino floor"~\cite{Monroe:2007xp}.} to experimental searches for new physics in the semi-dark $h\to Z X$ ($X \to \slashed{E}_T$) 
channel, below which a possible BSM signal would be buried. It also provides a target sensitivity for the $h \to Z X$ ($X \to \slashed{E}_T$) search at the LHC and future colliders which would guarantee a detection (albeit in that case not of BSM physics!), given by the SM branching fraction BR$(h\rightarrow Z\nu\Bar{\nu})_{\rm SM} = 5.4\cdot 10^{-3}$~\cite{deFlorian:2016spz}. 

\vspace{2mm}

\noindent {\bf LHC searches for $h\to Z X \to \ell\ell + \slashed{E}_T$.} We consider for the rest of this work an LHC center-of-mass energy $\sqrt{s} = 14$ TeV. 
Our analysis reveals the convenience of focusing on the production of the Higgs boson at the LHC in association with a $Z$ boson, $p p \to Z h$: 
For gluon-fusion (ggF) and vector boson fusion (VBF) Higgs production channels, the Higgs is either produced on its own (ggF) or recoiling against jets (ggF, VBF). Since the phase space for the Higgs decay $h \to Z X$ is fairly small (as $(m_h - m_Z)/m_h \ll 1 $), an accurate $\slashed{E}_T$ reconstruction may be limited by the transverse momentum ($p_T$) resolution of the jets. In addition, the $\ell\ell + \slashed{E}_T + \mathrm{jets}$ final state has very large SM backgrounds, in particular reducible ones if the $\slashed{E}_T$ reconstruction is not perfect. Higgs production in association 
with an electroweak gauge boson, $p p \to W^{\pm} h$ and $p p \to Z h$, is thus better suited for the $h \to Z X$ ($X \to \slashed{E}_T$) exotic Higgs decay search at the LHC.
Yet, the leptonic decay of the $W$ boson in $p p \to W h$ adds $\slashed{E}_T$ to the final state, making it challenging to disentangle this contribution from the Higgs boson decay products. In addition, the LHC cross section for the dominant SM background in this case, $p p \to W^{\pm} Z$, is very large, $\mathcal{O}(50)$~pb. 
In contrast, for $p p \to Z h$ ($h \to Z + \slashed{E}_T$) the 
leptonic decay of both $Z$ bosons offers a sharp reconstruction of the two di-lepton resonances together with an accurate $\slashed{E}_T$ measurement, combined with SM backgrounds that can be efficiently suppressed or are very small to begin with, as we discuss in detail below.

%
\vspace{1mm}



\begin{figure}[t]
\begin{centering}
\includegraphics[width=0.48\textwidth]{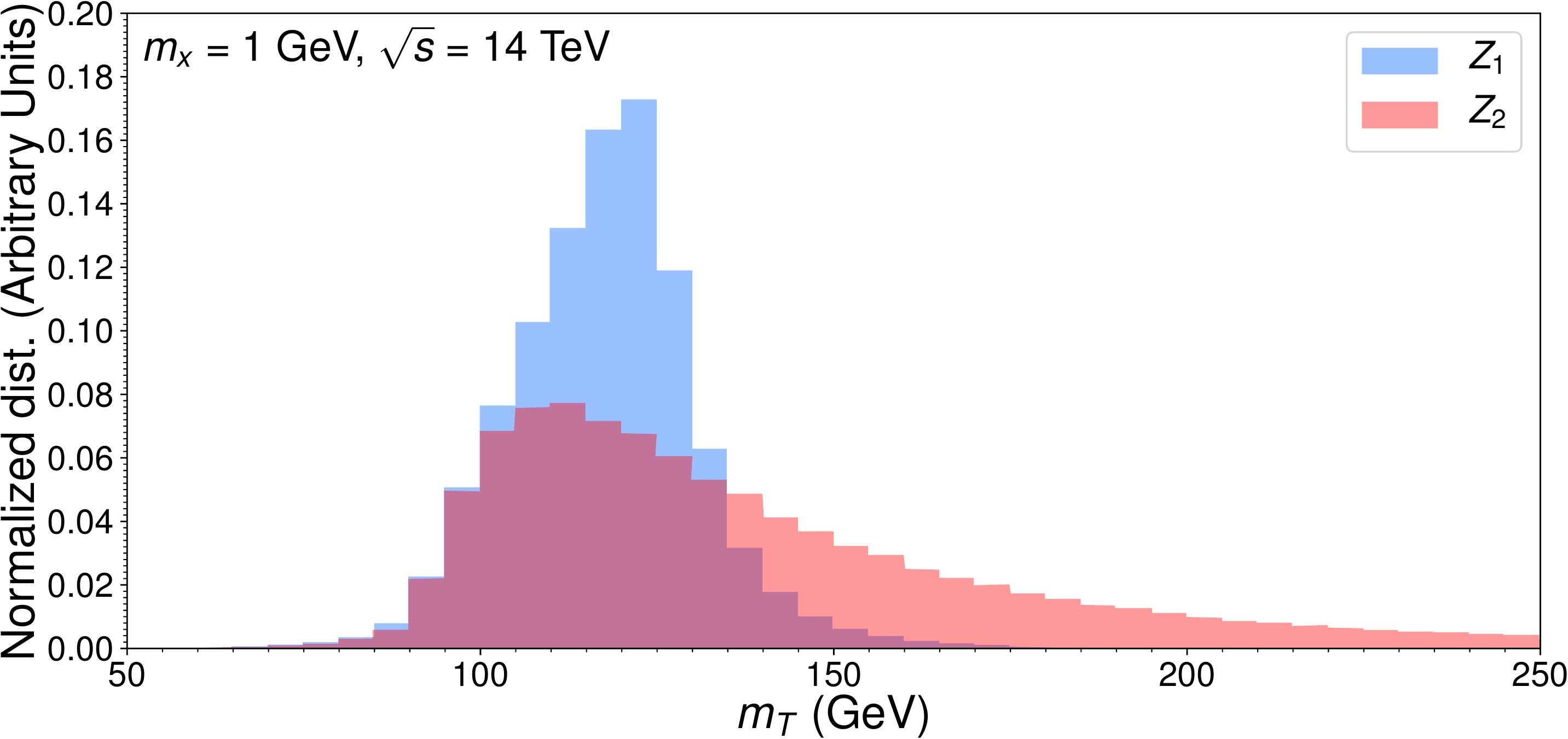} \\
\includegraphics[width=0.48\textwidth]{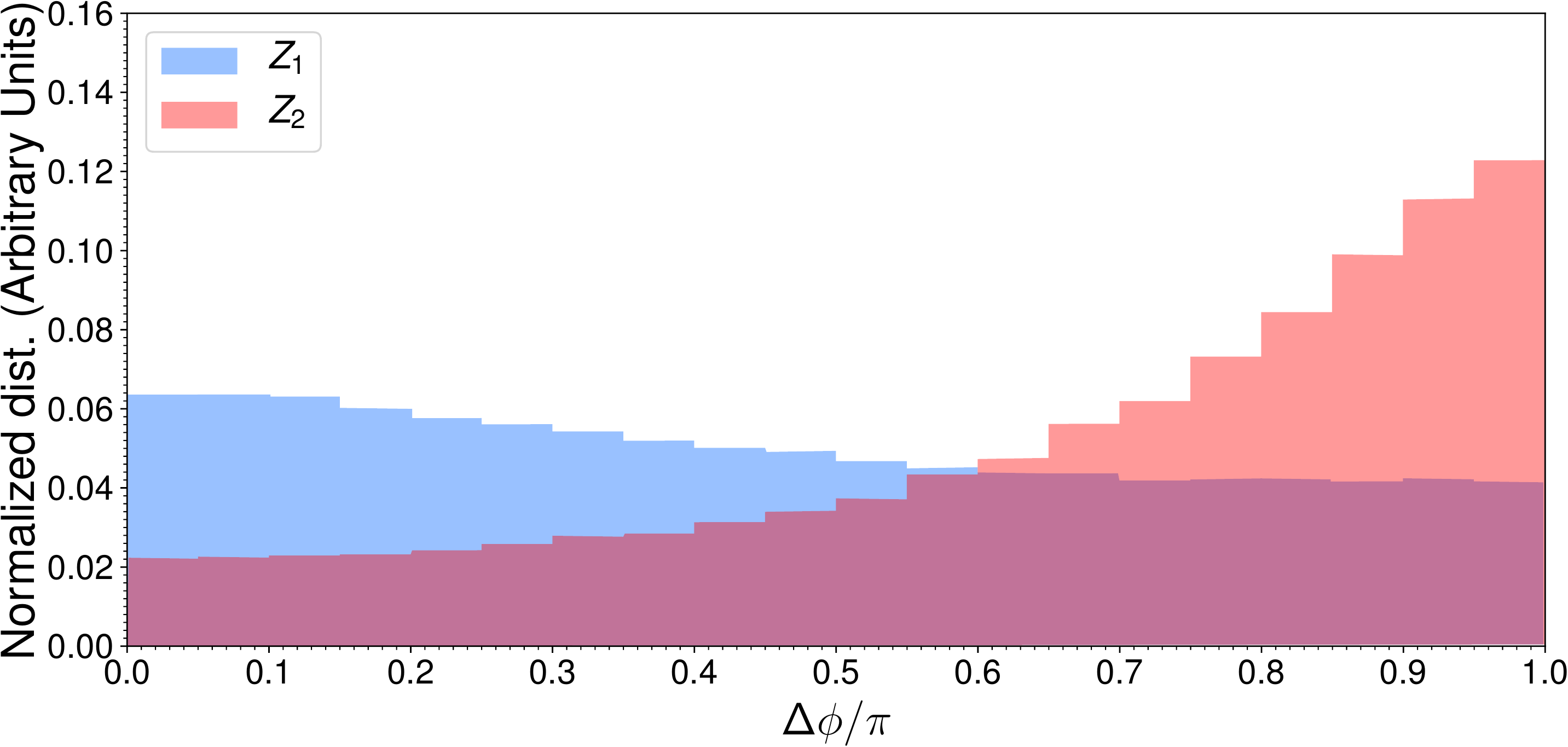} 

\caption{$M_T$ (top) and $\Delta \phi (Z,\vec \met)/\pi$ (bottom) for $Z_1$ (blue) and $Z_2$ (red), see text for details.}
\label{fig:Zselection}
\par\end{centering}

\vspace{-6mm}

\end{figure}

For our analysis, we generate the BSM signal specifically using a \texttt{FeynRules}~\cite{Alloul:2013bka} implementation of the two-Higgs-doublet-model plus pseudoscalar singlet (2HDM + $a$) extension of the SM (see e.g.~\cite{No:2015xqa,Goncalves:2016iyg,Bauer:2017ota}), through the decay $h \to Z a$ (with $a$ invisible). Our results apply to any two-body Higgs decay $h \to Z X$, $X \to \slashed{E}_T$. 
Both $Z$ bosons from the signal are considered to decay leptonically, $Z \to \ell \ell$. The relevant SM backgrounds are $p p \to Z Z \to 4 \ell$ (with $\slashed{E}_T$ appearing via mis-measurements and detector effects), $p p \to Z Z Z, W W Z \to 4 \ell + 2\nu$, $p p \to t\bar{t}Z, tWZ \to 4 \ell + 2\nu \,+ $ jets, and $p p \to Z h$ ($h \to W W^* \to 2\ell + 2 \nu$).
We generate signal and SM background event samples with {\sc MG5\_aMC@NLO 
}~\cite{Alwall:2014hca} (using the \textsc{NNPDF31$\_$nnlo}~\cite{Bertone:2017bme} parton distribution functions) with subsequent parton showering and hadronization via {\sc Pythia 8}~\cite{Sjostrand:2014zea} and detector simulation via {\sc Delphes 
}~\cite{deFavereau:2013fsa}, using the detector card designed for High Luminosity (HL)-LHC studies. 
%
We normalize the respective cross sections to their next-to-leading-order (NLO) in QCD values, obtained from the literature~\cite{2008:triboson, 2019:ttz} (for the $p p \to Z h$ and $p p \to Z Z$ processes the normalization is however performed to the NNLO cross section~\cite{deFlorian:2016spz, 2014:zz}; to avoid known issues at NLO in QCD related to real $b$-quark emission~\cite{Frixione:2008yi,Faham:2021zet}, $tWZ$ is kept at LO with a negligible impact on our analysis).
Selected events are required to contain exactly four reconstructed leptons after detector simulation, comprising two pairs of opposite-sign, same-flavor leptons. Events must pass the single, two or three-lepton trigger requirements from the ATLAS 2018 Trigger menu \cite{ATL-DAQ-PUB-2019-001}.
%
When multiple di-lepton combinations satisfying the selection requirements exist, the one minimizing $\Delta^2 = m_Z^{-2} [(m_{\ell\ell_1} - m_{Z})^2 + (m_{\ell\ell_2} - m_{Z})^2]$ (with $m_{\ell\ell_i}$ the di-lepton invariant masses) is chosen. Extra hadronic activity is vetoed by rejecting events with either $b$-tagged jets or hard jets with $p_T > 50$ GeV.


%


%


%

Since the Higgs decay is partially invisible, its invariant mass cannot be fully reconstructed, nor can the di-lepton pair from its decay be straightforwardly identified. The latter is key to better exploit the kinematic properties of the BSM signal in the analysis. We may identify the di-lepton pair corresponding to the $Z$ boson from the Higgs decay using the transverse mass $M_T$, 
given by
$M_T^2 = \left( \sqrt{m_{\ell\ell}^2 + |\Vec{p}_T^{\,\,\ell\ell}|^2} + \slashed{E}_T \right)^2 - \left| \Vec{p}_T^{\,\,\ell\ell} + \Vec{\slashed{E}}_T \right|^2$,
%
%
with $\vec \met$ and $\Vec{p}_T^{\,\,\ell\ell}$ the missing transverse 3-momentum and $Z$ boson transverse 3-momentum, respectively; a complementary approach would be to select the $Z$ boson closest to $\vec \met$ in the azimuthal plane as the one from the Higgs decay. Figure~\ref{fig:Zselection} shows the $M_T$ (top) and  
$\Delta \phi (Z,\vec \met)$ (azimuthal angle between $\vec \met$ and the 3-momentum of the di-lepton pair, bottom) distributions for the leptonically-decaying $Z$ boson from the Higgs decay (labelled as $Z_1$) and the $Z$ boson produced in association with the Higgs (labelled as $Z_2$). To optimally exploit the event kinematic information in identifying $Z_1$ and $Z_2$ for the BSM signal, we build a neural network (NN) (two hidden layers, 32 nodes each, using rectified linear unit activation for the hidden layers and a sigmoid function for the output) which takes as input $M_T$ and $\Delta \phi (Z,\vec \met)$ for both di-lepton pairs. The correct and wrong $Z_i$ assignments for the NN training are labelled using generator-level information. The NN is trained with a Monte Carlo sample of 20000 signal events (not used in our subsequent analysis) with $m_X = 1$ GeV, using the Adam algorithm for the optimisation. The efficiency obtained for a correct $Z_{1,2}$ choice for the signal is 73\%, and the NN is then applied in our sensitivity analysis to both the BSM signal (for $m_X \in [1,\,32.5]$ GeV) and the SM backgrounds.

%
\begin{figure}[h]
\begin{center}
\includegraphics[width=0.46\textwidth]{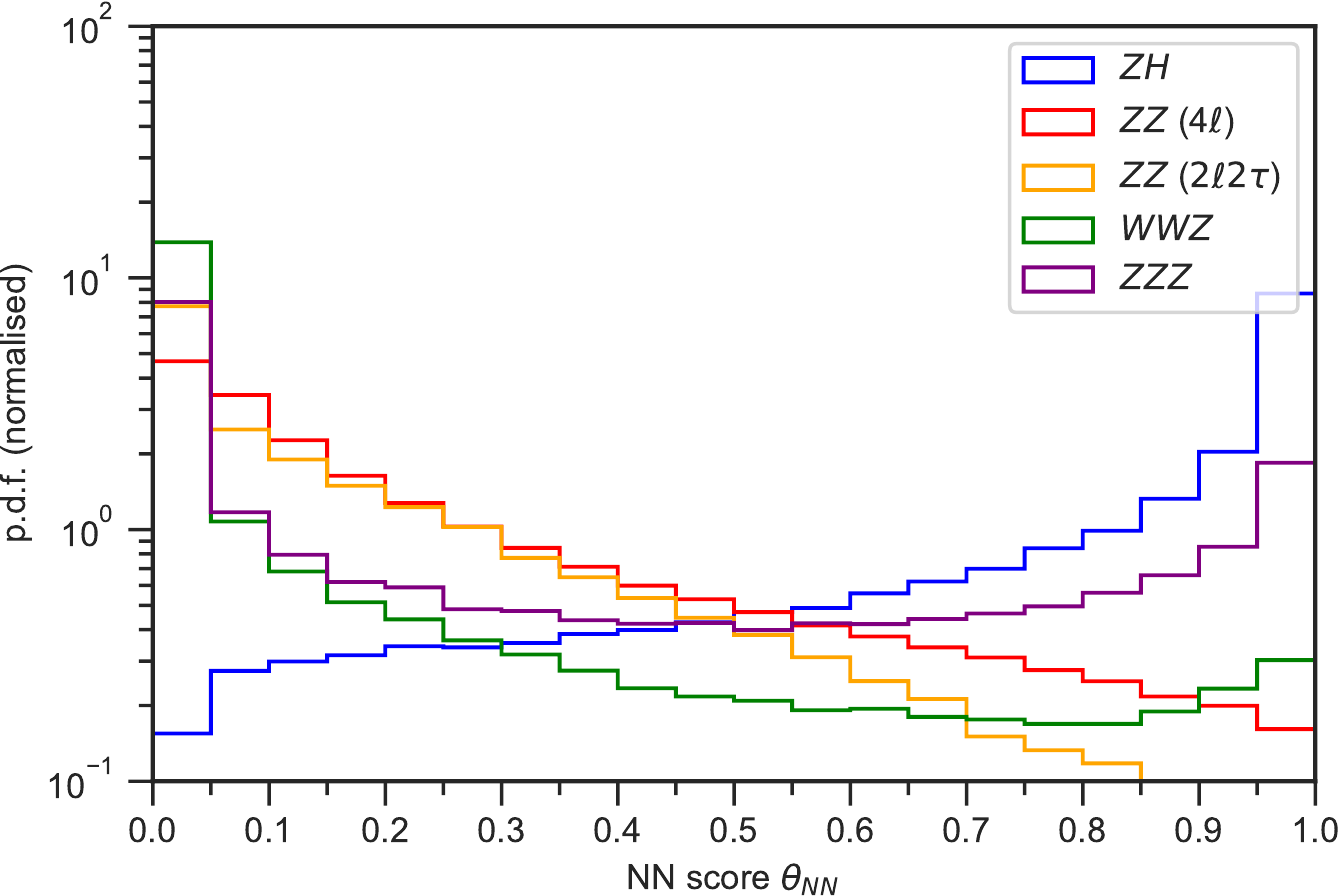} 
\caption{Score $\theta_{NN}$ of the neural network discriminating BSM signal vs SM background in our analysis, for the BSM signal with $m_X = 1$ GeV (labelled $ZH$, blue), and the relevant SM backgrounds: $ZZ \to 4\ell$ (red), $ZZ \to 2\ell 2\tau$ (yellow), $WWZ \to 4\ell + 2\nu$ (green), $ZZZ \to 4\ell + 2\nu$ (purple).}
\label{fig:NN}
\end{center}

\vspace{-6mm}

\end{figure}

The signal cross section (for BR$(h\rightarrow Z X) = 1$, BR$(X \rightarrow \slashed{E}_T) = 1$ and $m_X = 1$ GeV) after the initial event selection is 1.420 fb.
The respective SM background cross sections after event selection are 25.6 fb for $Z Z \to 4\ell$, 0.76 fb for $Z Z \to 2\ell + 2\tau$, 0.169 fb for $W W^{(*)} Z \to 4\ell + 2\nu$ (including the $p p \to Z h$, $h \to W W^*$ contribution), 0.012 fb for $Z Z Z \to 4\ell + 2\nu$, and 0.044 fb for $t\bar{t}Z,\, tWZ \to 4\ell + 2\nu\, +$ jets. 
Our $h \to Z X$ LHC signal region must target relatively high-$p_T$ $Zh$ associated production, with reconstructed $Z$-boson resonances for the two di-lepton pairs and the Higgs transverse mass $M_T$ from the $Z_1$ di-lepton candidate. Requiring a moderately large amount of $\met$, 
demanding $Z_1$ to be well-aligned with $\vec \met$ in the azimuthal plane and rejecting events with a large rapidity gap between di-lepton pairs also improves the sensitivity of the analysis.  
The rich event kinematics (four visible objects in the final state plus the missing transverse energy) indicates that a multivariate approach which accesses the 
full kinematic information of the events could enhance our BSM signal sensitivity.
We use another NN (two hidden layers of 256 nodes, same activation functions and optimisation as before) for the discrimination between signal and SM background, with input variables: $\met$, $m_{4\ell}$ (four-lepton invariant mass), $m_{\ell\ell_1}$ and $m_{\ell\ell_2}$, $\Delta \phi(Z_1,\vec \met)$ and $\Delta \phi(Z_2, \vec \met)$, $M_T$ (built from $Z_1$), $p_{T}^{\ell\ell_1}$ and $p_{T}^{\ell\ell_2}$ (di-lepton transverse momenta), $p_{T}^{\ell_1}$, $p_{T}^{\ell_2}$, $p_{T}^{\ell_3}$, $p_{T}^{\ell_4}$ (transverse momenta of the four leptons, ordered from higher to lower) and $(p_T^{\ell\ell_2}+\slashed{E}_T)/p_T^{\ell\ell_1}$. The NN is trained with an unbalanced Monte Carlo set dominated by $ZZ \to 4\ell$ events, precisely to optimise the rejection of this SM background (as it has by far the largest LHC cross section among SM backgrounds). 
%
%
The NN score $\theta_{NN}$ for the $m_X = 1$ GeV signal and relevant SM backgrounds is shown in Figure~\ref{fig:NN} (for $X$ of spin-1, angular correlations in the $Z_1$ di-lepton pair mildly differ from the spin-0 $X$ case analyzed here, yet our signal sensitivity results would be nearly unchanged). 

\begin{figure}[h]
\begin{center}
\includegraphics[width=0.48\textwidth]{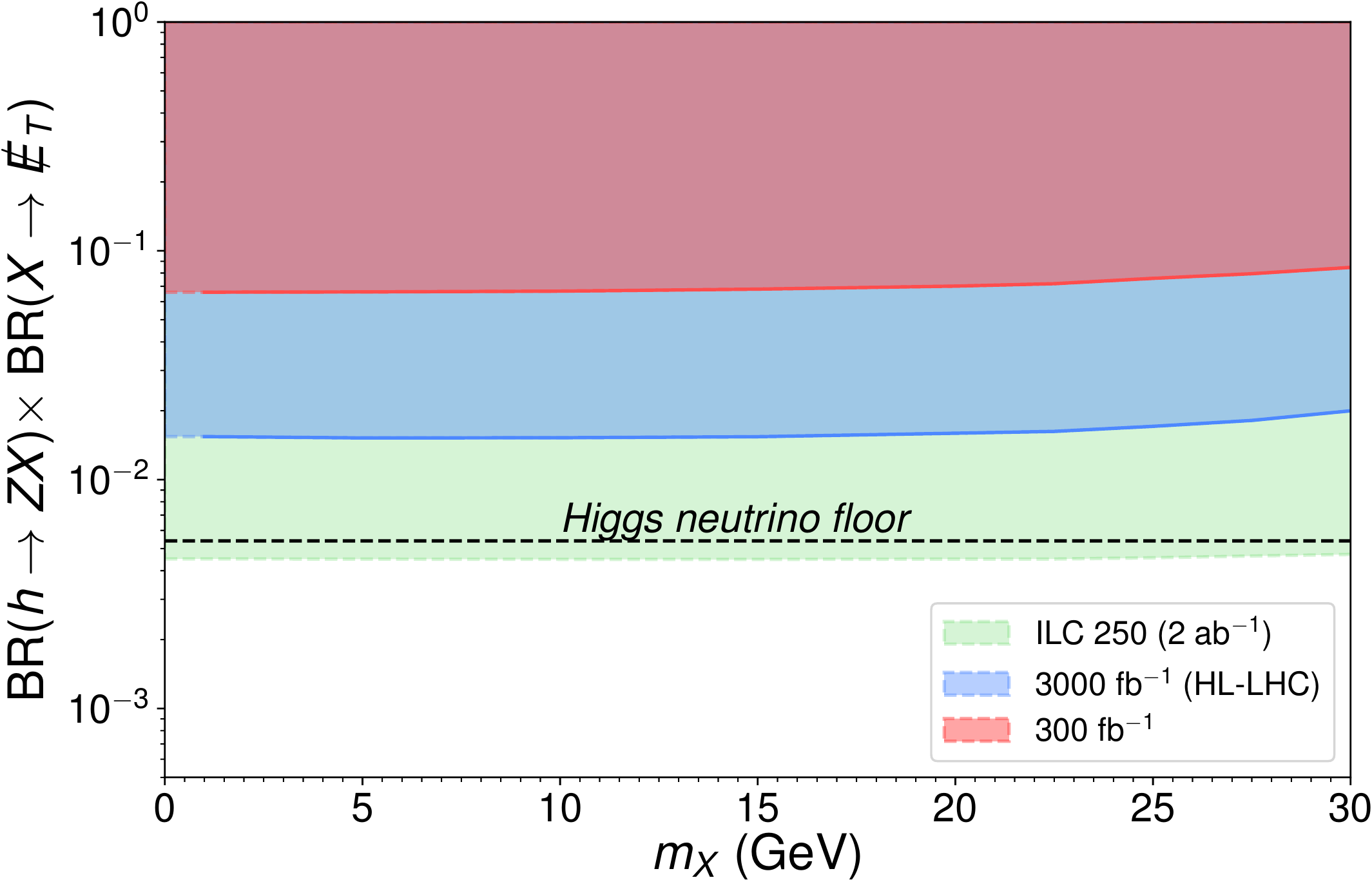}
\caption{$2\sigma$ exclusion sensitivity for BR$(h \to Z X) \times \rm{BR}(X\to \met)$ as a function of $m_X$ for an LHC integrated luminosity of 300 fb$^{-1}$ (red) and 3000 fb$^{-1}$ (HL-LHC, blue). The \textit{Higgs neutrino floor} is shown as a dashed-black line. The ILC $\sqrt{s} = 250$ GeV
(2 ab$^{-1}$) would-be sensitivity 
is shown in green.
}
\label{fig:BR}
\end{center}

\vspace{-6mm}

\end{figure}

Our signal region is defined for HL-LHC as $\theta_{NN} \geq 0.997$. The resulting signal and SM background efficiencies (evaluated on the NN test sets) are $0.12$, $1.5\cdot10^{-4}$, $2.8\cdot10^{-5}$, $0.0013$, $0.012$, $0.0016$ and $< 9.4 \cdot 10^{-4}$,  respectively for the signal (with $m_X = 1~\text{GeV}$), $ZZ \to 4\ell$, $WWZ$, $ZZ \to 2\ell + 2\tau$, $ZZZ$, $tWZ$ and $t\bar{t}Z$.
We employ the ``Asimov estimate'' \cite{2011Cowan} (since $\mathcal{O}(s/b)$ is not small) to derive the $2\sigma$ exclusion sensitivity on BR$(h\rightarrow Z X) \times \rm{BR}(X\to \met)$ with the $3$ ab$^{-1}$ of integrated luminosity from HL-LHC, in the range $m_X \in [1,\,32.5]$ GeV (our results may also be directly extrapolated to the $m_X \to 0$ limit). We find our NN results to improve by $\sim 30\%$ the sensitivity of a cut-and-count analysis, and come close to probing the \textit{Higgs neutrino floor} (for $m_X = 1$ GeV, it probes BR$(h\rightarrow Z X) \times \rm{BR}(X\to \met) = 2.8 \times \mathrm{BR}(h \to Z \nu \bar{\nu})_{\rm SM}$ at $2\sigma$). 
We repeat our analysis for an integrated luminosity of 300 fb$^{-1}$, with a less stringent signal region cut $\theta_{NN} \geq 0.985$ to increase the fraction of signal events surviving the selection. The sensitivity results for 300 fb$^{-1}$ and 3 ab$^{-1}$ (HL-LHC) are shown in Figure~\ref{fig:BR}.
%

\vspace{-2mm}

\noindent {\bf ILC searches for $h\to Z X \to \ell\ell + \slashed{E}_T$.}~A future International Linear Collider (ILC)~\cite{Behnke:2013xla,bambade2019international} operating at $\sqrt{s} = 250$ GeV 
would be able to probe BR$(h \to Z X)$ down to the \textit{Higgs neutrino floor} by exploiting several advantages over the LHC search discussed in the previous section: \textit{(i)} Higgstrahlung $e^+ e^- \to Z h$ is now the dominant Higgs production mode.~\textit{(ii)} The $e^{+} e^{-}$ collisions at ILC offer a much cleaner environment (largely void of hadronic activity) and the 3-momenta of the incoming particles is known up to radiative and smearing effects, allowing for full missing momentum reconstruction. \textit{(iii)} The Higgs recoil mass, constructed from the $Z$-boson recoiling against the Higgs boson ($Z_2$) as $M_{\rm reco}^2 = s + m_{Z_2}^2 - 2 E_{Z_2} \sqrt{s}$,
%
%
provides a straightforward way to correctly identify $Z_{1,2}$ for the BSM signal ($M_{\rm reco}$ built out of the $Z$-boson from the Higgs decay, $Z_1$, will not present any resonant structure).


For our analysis, we specifically consider $\sqrt{s} = 250$ GeV with 2 ab$^{-1}$ of integrated luminosity (90\% of it evenly split between the two opposite beam helicities) and beam polarizations being 80\% for the electrons and 30\% for the positrons respectively \cite{bambade2019international}.
Again, we consider the SM Higgs produced in association with a $Z$-boson, 
$e^+ e^- \to Z h$ for our BSM signal. The relevant SM backgrounds are now $e^+ e^- \to Z Z$ ($\to 4\ell,\,2\ell + 2\tau$), $e^+ e^- \to W W Z$ and $e^+ e^- \to Z Z \nu \bar{\nu}$ (including VBF initiated contributions; Higgs mediated contributions correspond to the SM \textit{Higgs neutrino floor}, and are not included) 
Our signal and background event generation is performed as in the previous section, using in this case the {\sc Delphes} detector card designed for ILC studies~\cite{deFavereau:2013fsa,Potter:2016pgp} (a study of lepton collider capabilities including the effects of initial state radiation or beamstrahlung is left for future work).

Our initial event selection mimics that of the previous LHC analysis. The cross sections for the signal (for BR$(h\rightarrow Z X) = 1$ and $m_X = 1$ GeV) and SM backgrounds after event selection are respectively 1.421 fb, 
5.64 fb (for $Z Z \to 4\ell$), 0.13 fb (for $Z Z \to 2\ell\, 2\tau$), 0.073 fb (for $W W^{(*)} Z \to 4\ell + 2\nu$, dominated by the $e^-e^+ \to Z h$, $h \to W W^*$ contribution), and 0.011 ab (for $Z Z \nu \bar{\nu} \to 4\ell + 2\nu$). 
For the ILC environment, the use of a NN does not offer such a strong advantage over a simpler (cut-and-count) analysis. 
We thus define our kinematic region for signal extraction in the latter way: we demand reconstructed $Z$-boson resonances for the two di-lepton pairs, $m_{Z_1} \in [55, 100]$ GeV, $m_{Z_2} \in [80, 105]$ GeV; we require the recoil mass constructed out of $Z_2$ in the range $M_{\rm reco} \in [120, 135]$ GeV, together with $p_{Z_2} > 50$ GeV and $\slashed{E} \in [5, 60]$ GeV ($p_{Z_2}$, $\slashed{E}$ respectively the modulus of the $Z_2$ di-lepton candidate's 3-momentum and the modulus of the missing 3-momentum);  the invariant mass $m_{Z_1}^{\rm miss}$ built from $Z_1$ and $\Vec{\slashed{E}}$, given by $(m_{Z_1}^{\rm miss})^2 = \left( \sqrt{m_{Z_1}^2 + p_{Z_1}^2} + \slashed{E} \right)^2 - \left| \Vec{p}_{Z_1} + \Vec{\slashed{E}} \right|^2$, is required to reconstruct the 125 GeV Higgs mass, $m_{Z_1}^{\rm miss} \in [95, 130]$ GeV;  we further require $m_{4\ell} > 160$ GeV, $(p_{Z_1}+\slashed{E})/p_{Z_2} < 1.8$ and $(m_{Z_2}^{\rm miss})^2 = \left( \sqrt{m_{Z_2}^2 + p_{Z_2}^2} + \slashed{E} \right)^2 - \left| \Vec{p}_{Z_2} + \Vec{\slashed{E}} \right|^2 > (95$ GeV$)^2$. These signal region cuts have an efficiency of $0.89$ for the BSM signal (for $m_X = 1$ GeV), and $1.7\cdot10^{-5}$, $0.013$, $0.085$, $0.24$ for the $ZZ\to 4\ell$, $ZZ\to 2\ell\, 2\tau$,  $WWZ$ and $ZZ\nu\bar{\nu}$ SM backgrounds, respectively. Using the ``Asimov estimate'', we  derive a $2\sigma$ sensitivity 
BR$(h\rightarrow Z X) \times \rm{BR}(X \to \slashed{E})= 0.0045$.
We show the corresponding ILC sensitivity as a function of $m_X$ in Figure~\ref{fig:BR}. We note that this sensitivity lies below the SM \textit{Higgs neutrino floor}, not been included in our analysis. This means that we should now instead consider the ILC discovery potential of BR$(h \to Z \nu\bar{\nu})_{\rm SM}$; the expected significance over the background only hypothesis reaches $2.4\sigma$ (the significance may be enhanced to $\sim 4\sigma$, at the expense of our BSM analysis not yielding a uniform sensitivity in $m_X$). 
ILC can thus sweep the entire new physics parameter space of semi-dark Higgs decays down to the~\textit{Higgs neutrino floor}. 

\vspace{2mm}

\noindent {\bf Constraints on specific models.} 

\vspace{1mm}

\noindent {\it (i) Axion-like particles.} The existence of interactions between the SM Higgs and light axion-like particles (ALPs) is a well-motivated BSM possibility~\cite{Brivio:2017ije,Bauer:2017ris}. Exotic Higgs decays represent a key experimental signature in this case. 
The Higgs boson partial decay width into a SM $Z$-boson and the ALP $a$ is $\Gamma(h\rightarrow Za) = (m_h^3/16\pi f_a^2)\, c_{aZh}^2 \, \lambda^{3/2}$, with
%
%
$f_a$ the ALP decay constant, $\lambda = (1- (m_Z^2-m_a^2)/m_h^2)^2 - 4 \, m_Z^2m_a^2/m_h^4 $ 
and $c_{aZh}$ the Wilson coefficient for the effective operator that couples the ALP to the SM Higgs field (which we leave here unspecified, see~\cite{Brivio:2017ije,Bauer:2017ris} for details). 
If $a$ then couples to some hidden sector particle(s) (see e.g.~\cite{Dolan:2017osp,Alves:2019xpc}), 
its dominant decay mode(s) may be into the dark sector, thus invisible at colliders. This encompasses the intriguing possibility that the ALP may be a mediator between the SM and the DM candidate~\cite{Dolan:2017osp}. 
%
Higgs decays $h\to Z a$, $a \to \met$ can then probe such ALP scenarios. 
%


\begin{figure}[h]
\begin{center}
\includegraphics[width=0.49\textwidth]{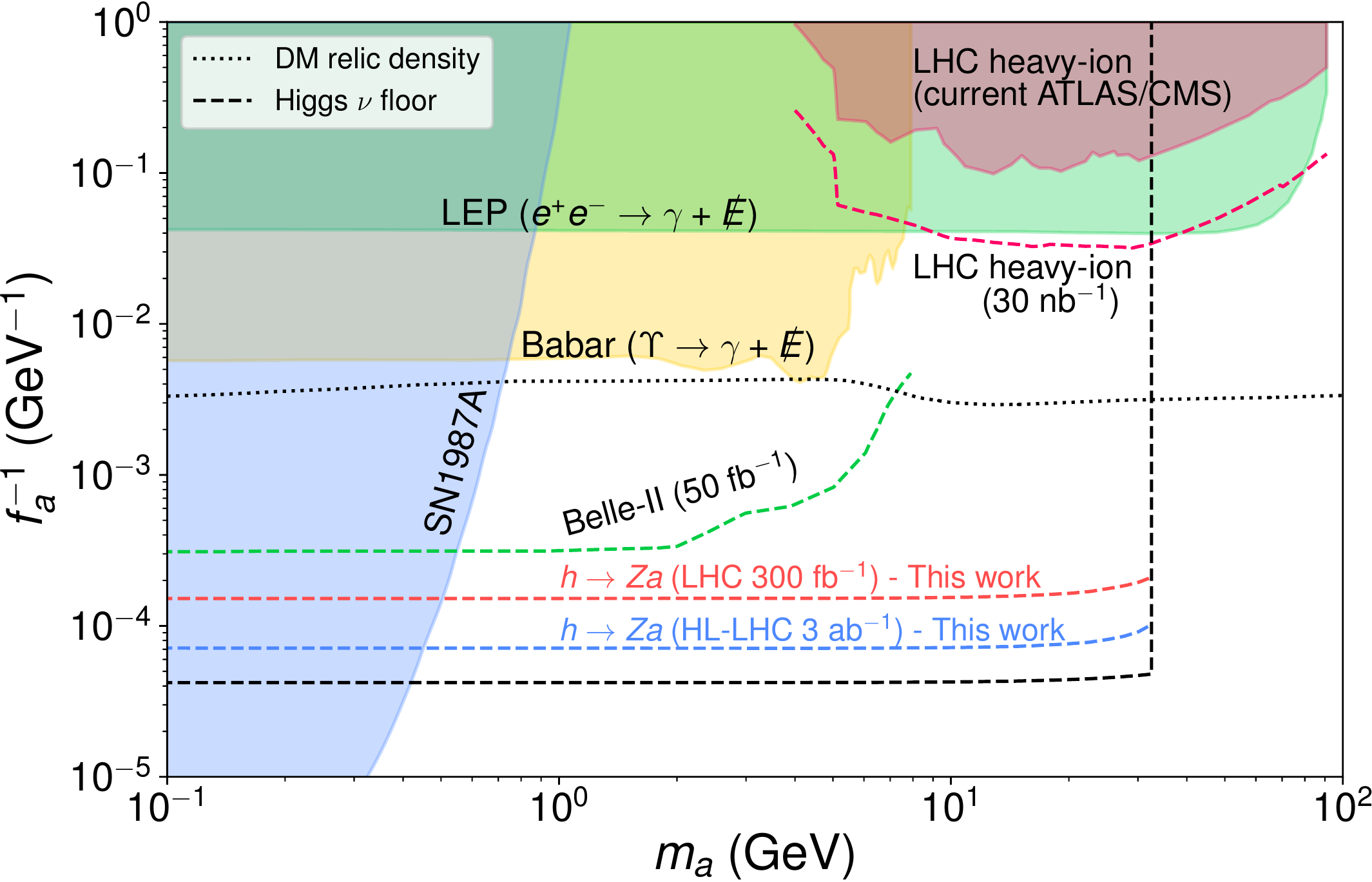}
\vspace{-4mm}
\caption{Present (solid) and projected (dashed) constraints on the ($m_a,\,f_a$) plane for an ALP with coupling to photons, a hidden (DM) fermion $\chi$ and the SM Higgs (via a $ c_{aZh}$ coupling), see text for details.}
\label{fig:ALPs}
\end{center}

\vspace{-6mm}

\end{figure}

To translate the model-independent LHC and ILC projected sensitivities from the previous sections into a probe of the parameter space of ALPs, we specifically consider, together with the coupling between the ALP and the SM Higgs, an ALP coupling to a hidden fermion $\chi$, given by $y_\chi\, \bar{\chi} \gamma^{\mu} \gamma^5 \chi\, \partial_{\mu} a/f_a$ as well as an ALP coupling to photons $c_{a\gamma\gamma}/f_a \, a\, F^{\mu\nu} \tilde{F}_{\mu\nu}$
(we do not include an ALP coupling to gluons or SM fermions for simplicity). 
$y_{\chi}$ does not have a preferred value, while the expectation for the bosonic Wilson coefficient is 
$c_{a\gamma\gamma} \sim \alpha_{\rm EM}$ (the electromagnetic coupling constant)~\cite{Alonso-Alvarez:2018irt}.
We then set $c_{a\gamma\gamma} = \alpha_{\rm EM}(Q)$ ($Q$ being the energy scale of the process considered), and $y_{\chi} = 1$, $c_{aZh} = 1$, $m_{\chi} = 0.45 \,m_a$ (to allow for the invisible ALP decay $a \to \chi\bar{\chi}$), and show in Figure~\ref{fig:ALPs} the LHC projected $2\sigma$ sensitivity on BR$(h\rightarrow Z a) \times \rm{BR}(a \to \chi\bar{\chi})$ in the ($m_a,\,f_a$) plane. We also depict the \textit{Higgs neutrino floor}, within reach of the $\sqrt{s} = 250$ GeV ILC. Figure~\ref{fig:ALPs} also shows, under the assumption that $\chi$ is the DM particle, the ($m_a,\,f_a$) relation yielding (for the choice of parameters described above) the observed DM relic abundance $\Omega_{\rm DM} h^2 = 0.12$~\cite{Planck:2015fie} generated via thermal freeze-out in the early Universe (taken from~\cite{Dolan:2017osp}), 
as well as the existing and projected constraints on this ALP scenario from searches at LEP, LHC and flavor factories (Babar, Belle-II), and astrophysics (supernova 1987A), all detailed in Appendix~I.  Finally, we also show in Appendix~I the corresponding limits on the ($m_a,\,f_a$) plane if a hypercharge coupling 
$c_{aBB}/f_a \, a\, B^{\mu\nu} \tilde{B}_{\mu\nu}$ (rather than a coupling only to photons) is assumed for the ALP.


\vspace{2mm}

\noindent {\it (ii) 2HDM + $a$.}~Two Higgs doublet models (2HDM) extended by a singlet pseudoscalar mediator and a fermionic singlet DM particle constitute the minimal renormalizable realization of a pseudoscalar portal to DM \cite{Ipek:2014gua,No:2015xqa,Goncalves:2016iyg,Bauer:2017ota,Robens:2021lov}, which avoids the stringent existing DM direct detection constraints~\cite{XENON:2018voc} (it yields a spin-independent DM-nucleon scattering cross section only at loop level~\cite{Ertas:2019dew,Abe:2019wjw}); it can also fit the observed gamma-ray galactic center excess~\cite{Ipek:2014gua,Tunney:2017yfp}; and it is a leading benchmark scenario for the DM interpretation of LHC searches~\cite{LHCDarkMatterWorkingGroup:2018ufk}). 
The potential of the 2HDM$ + \,a$ is~\cite{Bauer:2017ota}
\begin{eqnarray}
\label{Doublet_Singlet_potential} 
 V &=& V_{\rm 2HDM} + \frac{\mu^2_{a_0}}{2}\,a_0^2 + m_{\chi}\, \bar{\chi}\chi + \frac{\lambda_a}{4} a_0^4 + \lambda_{a1}\, a_0^2 \left|H_1\right|^2\nonumber\\ 
  &+& \lambda_{a2}\, a_0^2 \left|H_2\right|^2 
  + i\,\kappa\,a_0 \,H_1^{\dagger}H_2 + y_{\chi}\,a_0 \,\bar{\chi} i\gamma^{5} \chi 
 +   \mathrm{h.c.} 
\end{eqnarray}
%
with real pseudoscalar mediator $a_0$ and Dirac fermion DM $\chi$ with mass $m_{\chi}$, both singlets under the SM gauge interactions. $V_{\rm 2HDM}$ is the 2HDM scalar potential (with a $\mathbb{Z}_2$-symmetry softly-broken by a $\mu^2_{12} H_1^{\dagger}H_2 + h.c.$ term in $V_{\rm 2HDM}$, see e.g.~\cite{Gunion:2002zf}). The $\kappa$ term in \eqref{Doublet_Singlet_potential} yields a mixing $\theta$ between $a_0$ and the 2HDM neutral pseudoscalar state $A_0$, resulting in two mass eigenstates $a,\,A$ (with $m_a < m_A$) which provide the portal between the SM and DM.  
%

\begin{figure}[h]
\begin{center}
\includegraphics[width=0.49\textwidth]{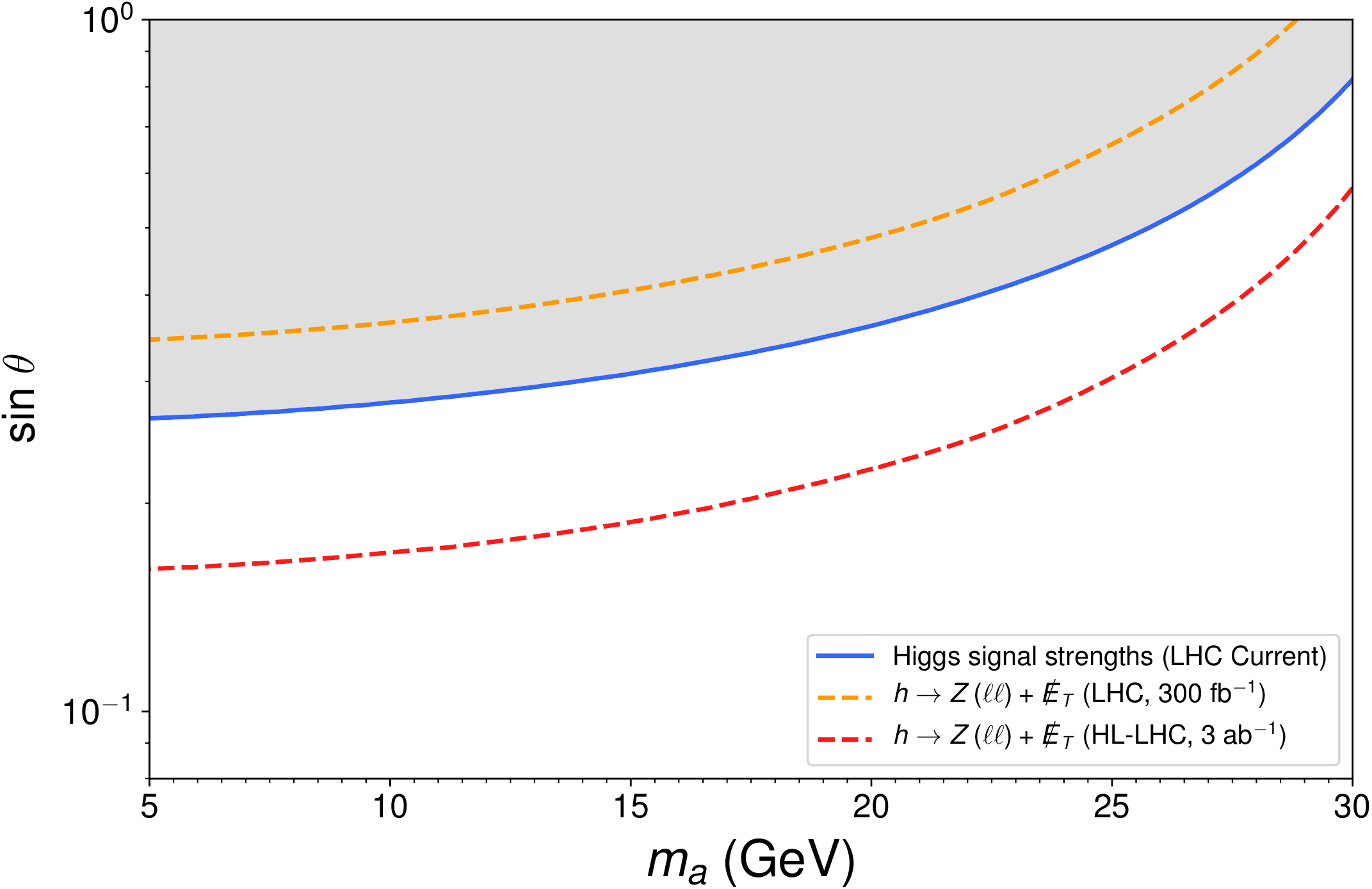}
\vspace{-4mm}
\caption{Present (solid, grey) and projected (dashed) constraints on the ($m_{a},\,\mathrm{sin} \, \theta$) plane for the 2HDM$ + \,a$ scenario analyzed in this work (with $\Gamma(h \to a a) = 0$), see text for details.}
\label{fig:2HDMa}
\end{center}

\vspace{-7mm}

\end{figure}

For a light pseudoscalar $a$ (singlet-like), the coupling between 
$a$, $h$ and $Z$  
leads to semi-dark Higgs decays ($a$ decays to DM particles with a branching fraction BR$(a\rightarrow\chi\Bar{\chi})\simeq 1$ unless $y_{\chi} \ll 1$). The partial decay width is $\Gamma(h\rightarrow Za)=(m_h^3/16\pi v^2)   \, c_{\beta-\alpha}^2 \,\sin^2{\theta}\,\lambda^{3/2}$, with $v$ the electroweak vev and $c_{\beta-\alpha} \equiv \cos (\beta - \alpha)$ parametrizing the deviation from the 2HDM alignment limit~\cite{Gunion:2002zf} (with $\tan{\beta}=v_2/v_1$, vev ratio of 2HDM Higgs doublets, and $\alpha$ the mixing angle between the 2HDM CP-even weak eigenstates~\cite{Gunion:1989we,Branco:2011iw}). 
The model also features an $h \to a a$ decay, leading to a Higgs invisible partial width via $a \to \chi\bar{\chi}$ decays. 
For $\left|c_{\beta-\alpha}\right| \ll 1$, as needed to satisfy the present LHC Higgs signal strength measurements \cite{CMS:2020gsy,ATLAS:2020qdt}, we generally expect
$\Gamma(h \to a a) \gg \Gamma(h \to Z a)$, yet in certain (albeit small) regions of the 2HDM$ + \,a$ parameter space, the $h \to Z a$ semi-dark Higgs decay can provide stronger sensitivity than the $h \to a a$ invisible Higgs decay, in particular when the $h - a - a$ coupling (see Appendix II for details) vanishes.  
We consider here a Type-I 2HDM (see~\cite{Branco:2011iw}) with $c_{\beta-\alpha} = 0.2$, $t_{\beta} \equiv \tan{\beta} = 6$, $M = 600$ GeV, $m_{H_0} = m_{H^{\pm}} = m_{A_0} =$ 700 GeV (with $M^2 = \mu_{12}^2/s_{\beta} c_{\beta}$ and with $H^{\pm}$ and $H_0$ the charged and neutral CP-even heavy 2HDM scalars). We choose in addition $m_{\chi} = 0.45 \, m_a$ and $y_{\chi}$ fixed to yield the observed DM relic density via thermal freeze-out (see e.g.~\cite{Goncalves:2016iyg,Tunney:2017yfp} and Appendix II). We further consider $\lambda_{a1} = \lambda_{a2}$ fixed in each case to the value that yields $\Gamma( h \to a a) = 0$, and show in Figure~\ref{fig:2HDMa} the projected LHC sensitivity (with 300 fb$^{-1}$ and at HL-LHC with 3 ab$^{-1}$) of the semi-dark Higgs decay $h \to Z (\ell\ell) + \met$ in the ($m_a$, sin $\theta$) parameter space of the 2HDM$ + \,a$. 
In addition, while our 2HDM ($c_{\beta-\alpha}$, $t_{\beta}$) benchmark satisfies both present and HL-LHC projected limits from Higgs signal strengths on 2HDM parameters~\cite{TheATLAScollaboration:2014qxe}, we find using the \textsc{ScannerS}~\cite{Muhlleitner:2020wwk} and \textsc{HiggsSignals}~\cite{Bechtle:2013xfa,Bechtle:2020uwn} numerical codes that $h \to Z a $ exotic Higgs decays in this scenario are currently constrained at $95\%$ C.L. to BR$(h \to Z a) < 0.042$ from Higgs signal strength measurements. This limit is also shown in Figure~\ref{fig:2HDMa}.   
Further details on these constraints, as well as a discussion on other searches that could be sensitive to the 2HDM$ + \,a$ parameter space (yet not to the specific benchmark we choose here) are given 
in Appendix II.  

\noindent {\it (iii) A comment on dark photons.} Light dark photons $Z_D$~\cite{Fabbrichesi:2020wbt} which interact with the SM via kinetic mixing (see e.g.~\cite{Jaeckel:2012yz}) give rise to an exotic Higgs decay $h \to Z Z_D$. Invisible dark photons would then constitute another new physics scenario that semi-dark Higgs decays could be sensitive to. However, current 95\% C.L. bounds on the kinetic mixing parameter $\epsilon$ from EW precision observables set 
$\epsilon < 0.03$ for dark photon masses $< 30$ GeV~\cite{Hook:2010tw}.
The corresponding $h \to Z Z_D$ branching fraction is then $< 10^{-3}$ (see e.g. \cite{Curtin:2013fra}), below the \textit{Higgs neutrino floor}.

\vspace{2mm}

\begin{center}
\textbf{Acknowledgements} 
\end{center}

\vspace{-1mm}

\begin{acknowledgements}
J.M.N. thanks Thomas Biekotter and Olalla Olea for assistance with the use of \textsc{ScannerS} and \textsc{HiggsSignals}.  J.A.A.S. acknowledges partial financial support by the Spanish ``Agencia Estatal de Investigaci\'on'' (AEI, MCIN/AEI/10.13039/501100011033) through the project PID2019-110058GB-C21. The work of J.M.C. was supported by the Spanish MICIU and the EU Fondo Social Europeo (FSE) through the grant PRE2018-083563.
The work of J.M.N. was supported by the Ram\'on y Cajal Fellowship contract RYC-2017-22986, and by grant PGC2018-096646-A-I00 from the Spanish Proyectos de I+D de Generaci\'on de Conocimiento.
J.M.N. also acknowledges support from the European Union's Horizon 2020 research and innovation programme under the Marie Sklodowska-Curie grant agreement 860881 (ITN HIDDeN), as well as from 
the AEI through the grant IFT Centro de Excelencia Severo Ochoa 
CEX2020-001007-S. 

\end{acknowledgements}


\subsection*{Appendix I: Details on ALP constraints} 

\vspace{-1mm}

We here discuss the constraints on the ($m_a,\,f_a$) parameter space of an invisibly-decaying ALP under the assumptions made in the main text. 
From~\cite{Dolan:2017osp}, we obtain the 95\% C.L. LEP limits from mono-photon searches~\cite{Fox:2011fx,DELPHI:2008uka}, which constrain an ALP produced via its coupling to photons and decaying invisibly, as well as the 90\% C.L. limits from $e^+ e^- \to \gamma + \slashed{E}$ (with $\slashed{E}$ the missing energy of the event) and rare upsilon decays into $\gamma + \slashed{E}$ from Babar~\cite{BaBar:2010eww,BaBar:2008aby} (see also~\cite{CrystalBall:1990xec}) and the projected 90\% C.L. sensitivity of Belle-II in the $\gamma + \slashed{E}$ final state. Also shown in Figure~\ref{fig:ALPs} are the current 95\% C.L. limits from heavy-ion (Pb-Pb) collisions at the LHC, from ALP searches in light-by-light scattering (as proposed in~\cite{Knapen:2016moh}) performed by ATLAS~\cite{ATLAS:2020hii} and CMS~\cite{CMS:2018erd} (see also~\cite{Bonilla:2022pxu}). We also include a projection drawn from rescaling the current ATLAS expected sensitivity to an (optimistic) integrated luminosity of 30 nb$^{-1}$. Finally, we show the bound from the energy loss of supernova 1987A from ALP emission, as taken from~\cite{Dolan:2017osp}. The supernova 1987A limit is stronger than usually quoted for an ALP coupling only to photons since the invisible decay of the ALP allows its corresponding energy to escape the supernova core even for parameter regions with a sizable coupling to photons (contrary to the usual case, where a large enough coupling to photons will result in the ALP being trapped in the core~\cite{Raffelt:2006cw}). 





We note that the existence of the invisible decay mode of the ALP leads (under the assumptions discussed in the main text) to a strongly suppressed ALP branching fraction to two photons, BR$(a \to \gamma\gamma) \sim 3\times 10^{-4}$. 
Limits from ALP searches in visible final states, like 
triphoton searches at LEP 1 and LEP 2 via the process $e^+ e^- \to \gamma^* \to \gamma \,a, \, a \to \gamma\gamma$ (as studied in~\cite{Mimasu:2014nea,Jaeckel:2015jla}), and searches for $Z \to \gamma \gamma$ decays\footnote{For light ALPs, with masses $\lesssim$ 10 GeV, the two photons from the ALP decay would appear merged in the detector, and
$e^+ e^- \to \gamma \gamma$ searches would be sensitive to the presence of the ALP~\cite{Jaeckel:2015jla}.} at LEP 1 have to be rescaled by BR$(a \to \gamma\gamma)$ (assumed to be 100\% in~\cite{Mimasu:2014nea,Jaeckel:2015jla}), and are too weak to appear in Figure~\ref{fig:ALPs}. Similarly, the dominant invisible decay of the ALP significantly weakens the limits from beam-dump experiments as compared to the case of visible
ALP scenarios (see e.g.~\cite{Dobrich:2015jyk}) roughly by a factor BR($a \to \gamma \gamma$) $\sim 10^{-4}$. A naive re-scaling of beam-dump limits results in no meaningful constraints (beyond what is currently excluded by other experiments/observations) from these, and we choose not to include them in Figure~\ref{fig:ALPs}. A precise re-derivation of these limits requires to additionally take into account the geometry of each experiment to obtain the expected number of $a \to \gamma\gamma$ events in the detector decay volume, which is beyond the scope of this paper.

\begin{figure}[h]
\begin{center}
\includegraphics[width=0.49\textwidth]{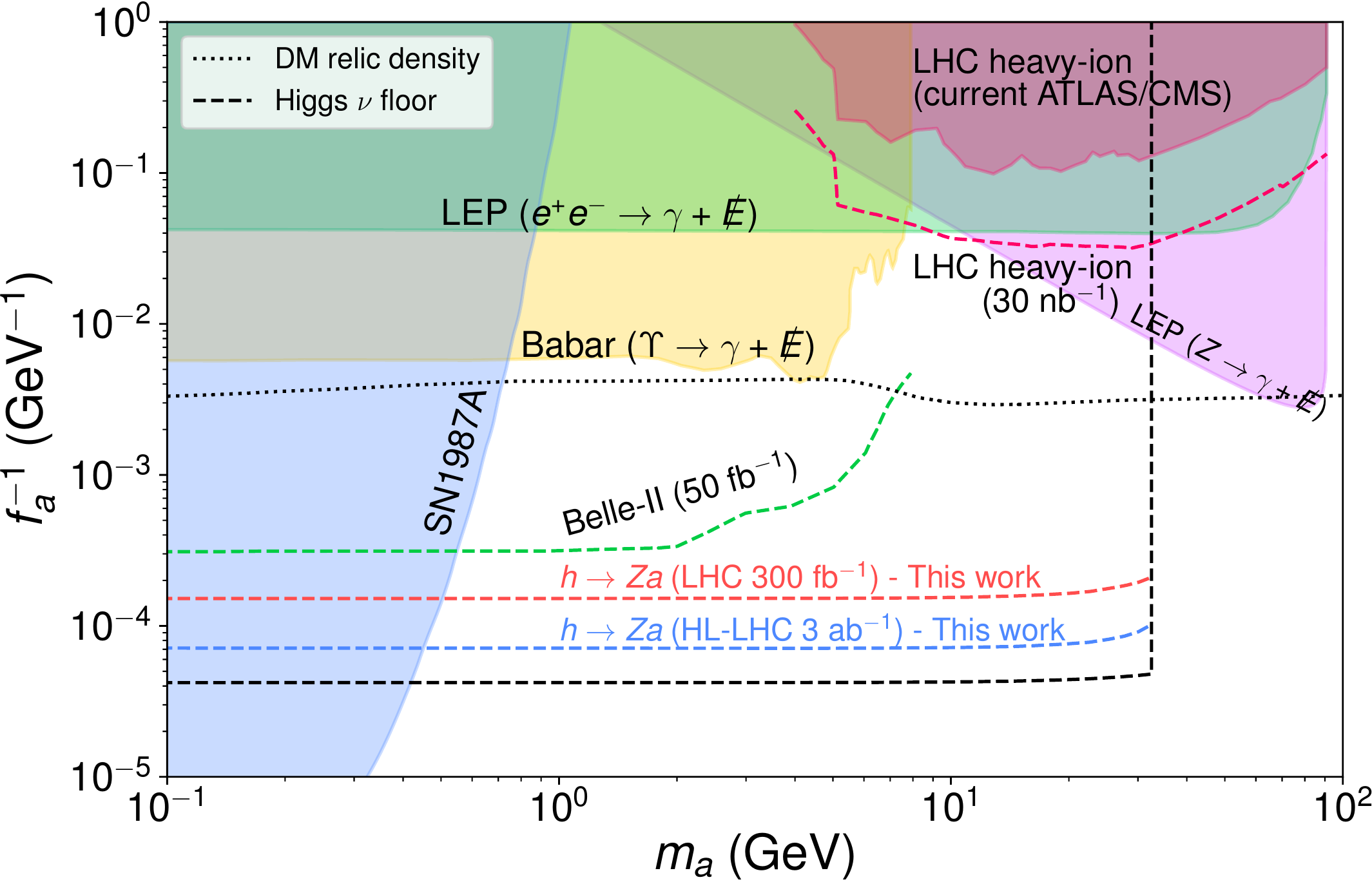}
\caption{Present (solid) and projected (dashed) constraints on the ($m_a,\,f_a$) plane for an ALP with coupling to the hypercharge field strength, a hidden (DM) fermion $\chi$ and the SM Higgs (via a $ c_{aZh}$ coupling), see text for details.}
\label{fig:ALPs_Hypercharge}
\end{center}

\vspace{-6mm}

\end{figure}

For an ALP coupled to the hypercharge field strength via $c_{aBB}/f_a \, a\, B^{\mu\nu} \tilde{B}_{\mu\nu}$, this introduces a coupling of the ALP to $Z Z$ and $Z \gamma$ (besides the already considered coupling to photons), given respectively by $c_{aZ Z}/f_a \, a\, Z^{\mu\nu} \tilde{Z}_{\mu\nu}$ and $c_{aZ\gamma}/f_a \, a\, F^{\mu\nu} \tilde{Z}_{\mu\nu}$, with $c_{aZZ} = \mathrm{sin}^2 \theta_W \, c_{aBB}$ and $c_{aZ\gamma} = -2\,\mathrm{sin}\, \theta_W \, \mathrm{cos} \,\theta_W \, c_{aBB}$ (with $\theta_W$ the weak mixing angle). Fixing $c_{aBB} = \alpha_{\rm EM}/\mathrm{cos}^2 \theta_W$ to match the ALP coupling to photons $c_{a\gamma\gamma} $ we assume in the main text, the above limits do not change, yet from $c_{aZ\gamma} \neq 0$ there is another constraint from LEP searches for rare $Z \to \gamma + a$ decays, with $a$ invisible. The L3 collaboration at LEP has set a limit BR$(Z \to \gamma + a) < 1.1 \times 10^{-6}$ at 90\% C.L.~\cite{L3:1997exg}, shown in Figure~\ref{fig:ALPs_Hypercharge} (in purple) together with the already considered constraints on our scenario (Figure~\ref{fig:ALPs}). Other potential bounds from rare $Z$ decays at LEP and LHC, e.g. from $Z \to 3\gamma$ or $Z \to \gamma \ell \ell$ (see~\cite{Bauer:2017ris}), do not lead to meaningful constraints in Figure~\ref{fig:ALPs_Hypercharge}.

\vspace{1mm}

Finally, we comment on the possibility of probing the ALP $a$ via exotic Higgs decays $h \to Z a$ with $a \to \gamma\gamma$, as discussed in~\cite{Bauer:2017ris}. We note that, while the corresponding final state allows to consider Higgs production in gluon-fusion (resulting in an~$\mathcal{O}(50)$ enhancement of the Higgs production cross section w.r.t. our scenario, which must rely on Higgs associated production), this is counteracted by the large suppression from BR$(a \to \gamma\gamma)$. A preliminary study of the LHC sensitivity to ALPs in exotic Higgs decays via $p p \to h \to Z a$, $Z\to \ell\ell$, $a \to \gamma\gamma$ including the leading SM backgrounds has been performed in~\cite{Brooijmans:2020yij}, indicating that such decay channel is much less sensitive than the one discussed in this work (given our assumptions for the ALP branching fractions).  


\subsection*{Appendix II: Details on 2HDM $+\, a$ constraints} 

\vspace{-2mm}

\begin{figure}[h]
\begin{center}
\includegraphics[width=0.49\textwidth]{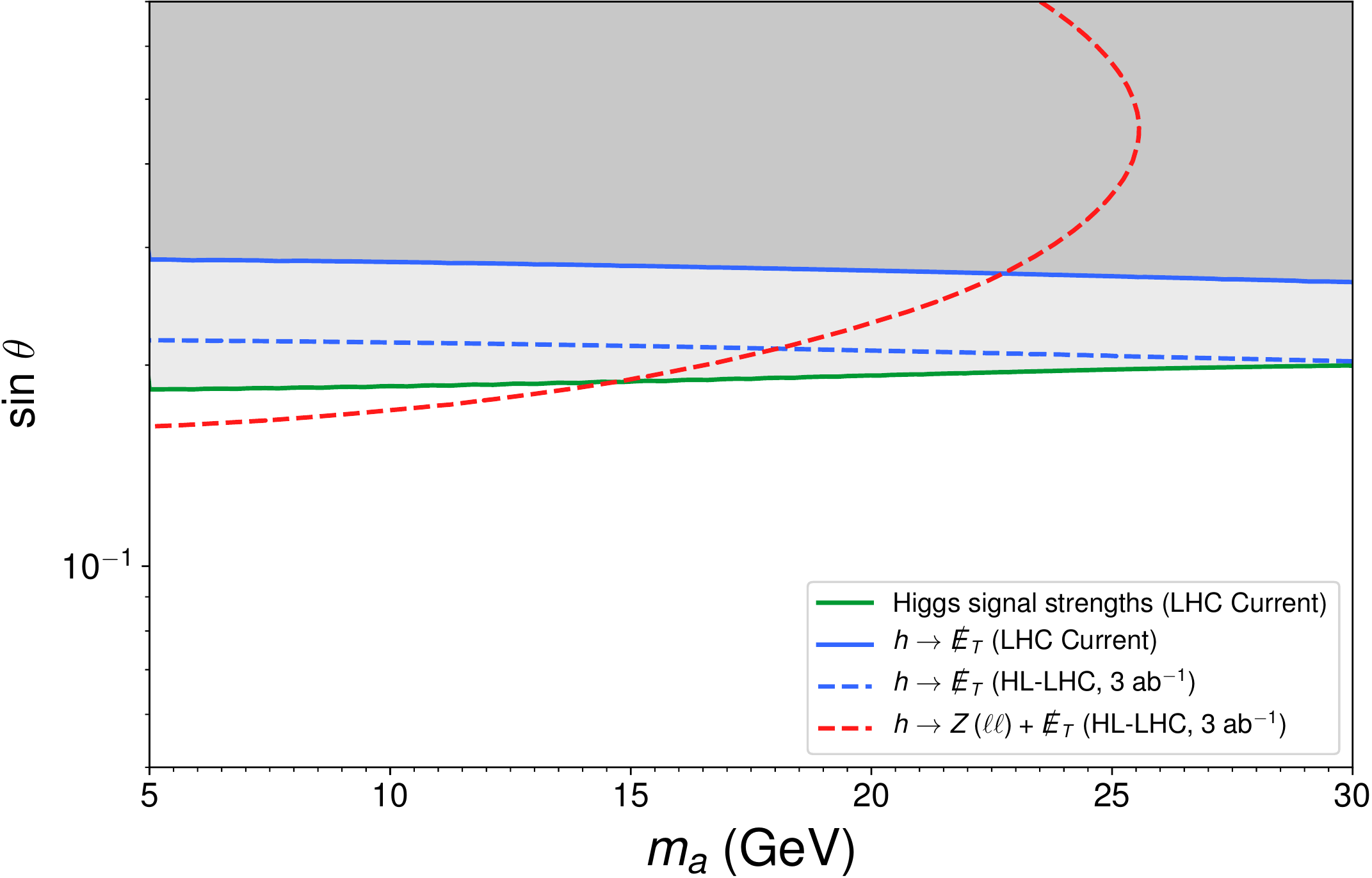}
\vspace{-4mm}
\caption{Present (solid, grey) and projected (dashed) constraints on the ($m_{a},\,\mathrm{sin} \, \theta$) plane for the 2HDM$ + \,a$ benchmark scenario analyzed in Appendix II (with $\Gamma (h \to a a) \neq 0$), see text for details.}
\label{fig:2HDMa_appendix}
\end{center}

\vspace{-5mm}

\end{figure}

The main direct experimental probes of the existence of a light ($m_a < 30$ GeV) singlet-like pseudoscalar $a$ in the 2HDM $+\, a$ are the exotic Higgs decays $h \to Z a$ (which this work explores in detail) and $h \to a a$. For the latter, the partial width is $\Gamma(h\rightarrow a a)=(v^2/32\,\pi\, m_h) \times g_{haa}^2\,\sqrt{1 - 4m_a^2/m_h^2}$, with the $h - a - a$ coupling $g_{haa}$ given in Eq.~(2). In the main text we have analyzed the tuned scenario $g_{haa} = 0$ (with varying $\lambda_{a1} = \lambda_{a2}$). We now consider the same 2HDM parameter benchmark as in the main text ($c_{\beta-\alpha} = 0.2$, $t_{\beta} = 6$, $M = 600$ GeV, $m_{H_0} = m_{H^{\pm}} = m_{A_0} =$ 700 GeV), but fix $\lambda_{a1} = \lambda_{a2} = 0$ such that $\Gamma(h \to a a) \neq 0$.  The resulting modified LHC sensitivity to the ($m_a$, sin $\theta$) parameter space of the 2HDM$ + \,a$ from  semi-dark Higgs decay $h \to Z (\ell\ell) + \met$ is shown in Figure~\ref{fig:2HDMa_appendix}. 
We also depict in Figure~\ref{fig:2HDMa_appendix} the constraint on the Higgs invisible width from $h \to a a$ decays, which at present is BR$(h\rightarrow \met)<0.11$ at $95\%$ C.L.~\cite{ATLAS:2020kdi} under the assumption of SM Higgs production, and is expected to be BR$(h\rightarrow \met)<0.04$ at $95\%$ C.L.~\cite{CMS:2018tip} at the HL-LHC.

\vspace{1mm}

In addition, LHC Higgs signal strengths constrain the 2HDM $+\, a$ parameter space for $c_{\beta-\alpha} \neq 0$ and/or $\Gamma(h \to Z a),\,\Gamma(h \to a a) \neq 0$. For $c_{\beta-\alpha} = 0.2$, $t_{\beta} = 6$, we have performed a global $\chi^2$ fit to present Higgs signal strength measurements via the  \textsc{HiggsSignals}~\cite{Bechtle:2013xfa,Bechtle:2020uwn} numerical code interfaced to \textsc{ScannerS}~\cite{Muhlleitner:2020wwk}, yielding the constraint BR$(h \to Z a) \,+$ BR$(h \to a a) < 0.042$ at 95\% C.L., shown in Figure~\ref{fig:2HDMa_appendix}. While this bound will certainly improve at the HL-LHC, the corresponding sensitivity improvement in sin$\,\theta$ will only be mild since $\Gamma(h \to Z a) \propto \mathrm{sin}^2\theta$ and $\Gamma(h \to a a) \propto \mathrm{sin}^4\theta$ (for our $\lambda_{a1} = \lambda_{a2} = 0$ benchmark), and we expect direct searches for $a$ in exotic Higgs decays to remain competitive with indirect probes through Higgs signal strength measurements.  

\begin{widetext}

\label{eq_g_haa}
\begin{eqnarray}
g_{haa} &=& \left[ \frac{2\, m_a^2}{v^2} s_{\beta-\alpha} + \frac{2 (m_{H_0}^2 - M^2) - m_h^2 + (m_{H_0}^2- m_h^2) [1 + s_{\beta-\alpha}(s_{\beta-\alpha} - c_{\beta-\alpha} (t_{\beta} - t_{\beta}^{-1})) ]}{v^2} s_{\beta-\alpha} - \lambda_7 c_{\beta-\alpha} \right] s_{\theta}^2 \nonumber   \\
&+& 2 \left( \frac{\lambda_{a1} + t_{\beta}^2 \lambda_{a2}}{1 + t_{\beta}^2} s_{\beta-\alpha} - \frac{(\lambda_{a2} - \lambda_{a1}) t_{\beta}}{1 + t_{\beta}^2} c_{\beta-\alpha}  \right) c_{\theta}^2\, , \\
\lambda_7 &=& \frac{(M^2 - m_{H_0}^2)(t_{\beta} - t_{\beta}^{-1}) - c_{\beta-\alpha} (m_{H_0}^2- m_h^2)  (s_{\beta-\alpha} - c_{\beta-\alpha} (t_{\beta} - t_{\beta}^{-1}))}{v^2} \nonumber
\end{eqnarray}


\end{widetext}

Other searches for the state $a$ do not provide meaningful sensitivity in the scenario we consider: searches for $h \to a a$ in visible final states (see~\cite{Cepeda:2021rql} for a review) like $b\bar{b} \tau\tau$~\cite{CMS:2018zvv} and $\tau\tau \tau\tau$~\cite{CMS:2019spf} are found to be $\mathcal{O}(10^3)$ less sensitive than probes of the Higgs invisible width, and fall short of providing any limit on BR($h \to a a$) by a factor $\sim 50-100$, with searches in other final states (e.g.~$b\bar{b} \mu\mu$~\cite{ATLAS:2021hbr}, $\tau\tau \mu\mu$~\cite{CMS:2018qvj,CMS:2020ffa}) yielding even smaller sensitivity. Such visible decays of $a$ are then generally not relevant in the 2HDM $+\, a$ with $m_{a} > 2\, m_{\chi}$, since 
matching the observed DM relic density requires $y_{\chi} \sim 1$ (see below), leading to BR$(a \to \chi \bar{\chi}) > 0.99$ in general.
We also find that current LHC mono-jet searches~\cite{CMS:2021snz} fall short of probing any region of the ($m_a$, sin $\theta$) plane of Figures~\ref{fig:2HDMa} and~\ref{fig:2HDMa_appendix} by a factor $\sim 1/(3 \,t_{\beta}^2)$.
Finally, we note that, while the $a-h-Z$ coupling could be constrained via Higgs boson production in association with missing energy at LEP2 through the process $e^+ e^- \to Z^* \to a h$,   
this does not yield meaningful limits on the 2HDM $+\, a$ parameter space: 
the searches for $h + \bar{\nu}\nu$ signatures by the OPAL~\cite{Abbiendi:2002yk}, L3~\cite{Achard:2001pj} and ALEPH~\cite{Heister:2001kr,Barate:2000ts} experiments at LEP impose a constraint on the missing mass $M_{\rm miss}$ of the event (equal to $m_a$ in our scenario) which is not fulfilled by our signal (e.g.~the OPAL analysis~\cite{Abbiendi:2002yk} requires $50$ GeV $< M_{\rm miss} < 130$ GeV, and it is thus not sensitive to $m_a < 30$ GeV). The corresponding search by 
the DELPHI experiment~\cite{Abdallah:2003ip}, while not imposing such a cut on $M_{\rm miss}$, does not consider Higgs boson masses above $120$ GeV.


We also consider possible constraints on the spin-0 states from the 2HDM, $H^{\pm}$, $H_0$ and $A$ (doublet-like). Electroweak precision observables (EWPO) constrain (dominantly via the oblique $T$-parameter)  the mass splittings among the 2HDM scalars, since the 2HDM scalar potential of the 2HDM is custodially invariant for $m_{A_0}=m_{H^\pm}$ or $m_{H_0}=m_{H^\pm}$. The latter is chosen for our benchmark scenario, directly satisfying EWPO. Potential constraints from flavour physics, in particular from $B$-meson decays: $\bar{B}\to X_s \gamma$ decays (which constrain $m_{H^{\pm}}$~\cite{Hermann:2012fc,Misiak:2020vlo}) and from $B_s\to\mu^+\mu^-$ (which constrain the presence of a light $a$ coupling to SM fermions~\cite{Skiba:1992mg,Logan:2000iv}), are also directly satisfied for a Type-I 2HDM with moderately high $t_{\beta}$ (we have chosen in particular $t_{\beta} = 6$ in our benchmark analysis). Finally, we also discuss direct searches for the 2HDM states as a probe of our scenario: away from the 2HDM alignment limit, $H_0 \to W^+ W^-$ and $H_0 \to Z Z$ decays could yield sensitivity if the mass scale of the 2HDM scalars is not very high. At the same time, 
$H_0 \to Z a$ decays would lead to resonant mono-$Z$ signatures~\cite{No:2015xqa,Bauer:2017ota}, which have been recently been searched for by the ATLAS experiment~\cite{ATLAS:2021gcn} (ATLAS also searches for this final state in $H_0 \to Z Z \to \ell\ell \nu \bar{\nu}$ decays~\cite{ATLAS:2020tlo}). However, in all these cases, we find that for $|s_{\beta-\alpha} t_{\beta}^{-1} - c_{\beta-\alpha}| \ll 1$ (as our scenario features) the production of $H_0$ at the LHC is suppressed, and no meaningful limit is obtained.

To conclude, we discuss the need to reproduce the observed DM relic density $\Omega_{\mathrm{DM}} h^2 = 0.12$~\cite{Planck:2015fie} within the 2HDM $+\, a$, via DM thermal freeze-out in the early-Universe. For $m_{\chi} \gtrsim 2$ GeV, the DM annihilation cross section into SM particles (via $\chi\bar\chi\to q_{\rm SM} q_{\rm SM}$, with $q_{\rm SM}$ here being generic SM particles) in the nonrelativistic limit is
\begin{align}
\langle \sigma \mathrm{v}\rangle&=\frac{y_\chi^2}{2\pi}\frac{m_\chi^2}{m_a^4\, t^2_\beta}s_{\theta}^2 c_{\theta}^2\,
\left[\left(1-\frac{4m_\chi^2}{m_a^2}\right)^2+\frac{\Gamma_a^2}{m_a^2}\right]^{-1}
\nonumber
\\
&\quad\quad\quad\quad\times\sum_{f} N_C \frac{m_f^2}{v^2}\sqrt{1-\frac{m_f^2}{m_a^2}}.
\label{eq:annih}
\end{align}
with $\Gamma_a$ the decay width of $a$. The sum over SM fermion $f$ annihilation channels involves quarks ($N_C=3$) and charged 
leptons ($N_C=1$). Reproducing the observed DM relic density via thermal freeze-out requires $\langle \sigma \mathrm{v}\rangle \simeq 3\times 10^{-26}{\rm  cm}^3/{\rm s}$, which generically leads to $\mathcal{O}(1)$ values for $y_{\chi}$. For $m_{\chi} < 2$ GeV the DM annihilation into SM fermions ($b,c$-quarks and/or $\tau$-leptons, depending on $m_{\chi}$) ceases to be the dominant DM annihilation process, and instead annihilation into QCD hadrons (via the 1-loop coupling of $a$ to gluons) dominates. Due to its complexity, we do not study that region of the 2HDM $+\, a$ parameter space, which may also involve $y_{\chi} \gg 1$, in this work.

\vspace{2mm}

\bibliographystyle{JHEP-cerdeno}
\bibliography{biblio}

\providecommand{\href}[2]{#2}\begingroup\raggedright\begin{thebibliography}{100}

\bibitem{Curtin:2013fra}
D.~Curtin et~al., \emph{{Exotic decays of the 125 GeV Higgs boson}},
  \href{http://dx.doi.org/10.1103/PhysRevD.90.075004}{\emph{Phys. Rev. D} {\bf
  90} (2014) 075004}, [\href{http://arxiv.org/abs/1312.4992}{{\tt 1312.4992}}].

\bibitem{CMS:2021pcy}
{\scshape CMS} collaboration, A.~Tumasyan et~al., \emph{{Search for low-mass
  dilepton resonances in Higgs boson decays to four-lepton final states in
  proton\textendash{}proton collisions at $\sqrt{s}=13\,\text {TeV} $}},
  \href{http://dx.doi.org/10.1140/epjc/s10052-022-10127-0}{\emph{Eur. Phys. J.
  C} {\bf 82} (2022) 290}, [\href{http://arxiv.org/abs/2111.01299}{{\tt
  2111.01299}}].

\bibitem{ATLAS:2021ldb}
{\scshape ATLAS} collaboration, G.~Aad et~al., \emph{{Search for Higgs bosons
  decaying into new spin-0 or spin-1 particles in four-lepton final states with
  the ATLAS detector with 139 fb$^{-1}$ of $pp$ collision data at $\sqrt{s}=13$
  TeV}},  \href{http://arxiv.org/abs/2110.13673}{{\tt 2110.13673}}.

\bibitem{CMS:2020ffa}
{\scshape CMS} collaboration, A.~M. Sirunyan et~al., \emph{{Search for a light
  pseudoscalar Higgs boson in the boosted $\mu\mu\tau\tau$ final state in
  proton-proton collisions at $\sqrt{s}=$ 13 TeV}},
  \href{http://dx.doi.org/10.1007/JHEP08(2020)139}{\emph{JHEP} {\bf 08} (2020)
  139}, [\href{http://arxiv.org/abs/2005.08694}{{\tt 2005.08694}}].

\bibitem{CMS:2018qvj}
{\scshape CMS} collaboration, A.~M. Sirunyan et~al., \emph{{Search for an
  exotic decay of the Higgs boson to a pair of light pseudoscalars in the final
  state of two muons and two $\tau$ leptons in proton-proton collisions at $
  \sqrt{s}=13 $ TeV}},
  \href{http://dx.doi.org/10.1007/JHEP11(2018)018}{\emph{JHEP} {\bf 11} (2018)
  018}, [\href{http://arxiv.org/abs/1805.04865}{{\tt 1805.04865}}].

\bibitem{CMS:2019spf}
{\scshape CMS} collaboration, A.~M. Sirunyan et~al., \emph{{Search for light
  pseudoscalar boson pairs produced from decays of the 125 GeV Higgs boson in
  final states with two muons and two nearby tracks in pp collisions at
  $\sqrt{s}=$ 13 TeV}},
  \href{http://dx.doi.org/10.1016/j.physletb.2019.135087}{\emph{Phys. Lett. B}
  {\bf 800} (2020) 135087}, [\href{http://arxiv.org/abs/1907.07235}{{\tt
  1907.07235}}].

\bibitem{ATLAS:2020ahi}
{\scshape ATLAS} collaboration, G.~Aad et~al., \emph{{Search for Higgs boson
  decays into two new low-mass spin-0 particles in the 4$b$ channel with the
  ATLAS detector using $pp$ collisions at $\sqrt{s}= 13$ TeV}},
  \href{http://dx.doi.org/10.1103/PhysRevD.102.112006}{\emph{Phys. Rev. D} {\bf
  102} (2020) 112006}, [\href{http://arxiv.org/abs/2005.12236}{{\tt
  2005.12236}}].

\bibitem{ATLAS:2021hbr}
{\scshape ATLAS} collaboration, G.~Aad et~al., \emph{{Search for Higgs boson
  decays into a pair of pseudoscalar particles in the $bb\mu\mu$ final state
  with the ATLAS detector in $pp$ collisions at $\sqrt{s}=13$ TeV}},
  \href{http://arxiv.org/abs/2110.00313}{{\tt 2110.00313}}.

\bibitem{CMS:2018nsh}
{\scshape CMS} collaboration, A.~M. Sirunyan et~al., \emph{{Search for an
  exotic decay of the Higgs boson to a pair of light pseudoscalars in the final
  state with two muons and two b quarks in pp collisions at 13 TeV}},
  \href{http://dx.doi.org/10.1016/j.physletb.2019.06.021}{\emph{Phys. Lett. B}
  {\bf 795} (2019) 398--423}, [\href{http://arxiv.org/abs/1812.06359}{{\tt
  1812.06359}}].

\bibitem{CMS:2018zvv}
{\scshape CMS} collaboration, A.~M. Sirunyan et~al., \emph{{Search for an
  exotic decay of the Higgs boson to a pair of light pseudoscalars in the final
  state with two b quarks and two $\tau$ leptons in proton-proton collisions at
  $\sqrt{s}=$ 13 TeV}},
  \href{http://dx.doi.org/10.1016/j.physletb.2018.08.057}{\emph{Phys. Lett. B}
  {\bf 785} (2018) 462}, [\href{http://arxiv.org/abs/1805.10191}{{\tt
  1805.10191}}].

\bibitem{ATLAS:2018pvw}
{\scshape ATLAS} collaboration, M.~Aaboud et~al., \emph{{Search for the Higgs
  boson produced in association with a vector boson and decaying into two
  spin-zero particles in the $H \rightarrow aa \rightarrow 4b$ channel in $pp$
  collisions at $\sqrt{s} = 13$ TeV with the ATLAS detector}},
  \href{http://dx.doi.org/10.1007/JHEP10(2018)031}{\emph{JHEP} {\bf 10} (2018)
  031}, [\href{http://arxiv.org/abs/1806.07355}{{\tt 1806.07355}}].

\bibitem{ATLAS:2018jnf}
{\scshape ATLAS} collaboration, M.~Aaboud et~al., \emph{{Search for Higgs boson
  decays into pairs of light (pseudo)scalar particles in the $\gamma\gamma jj$
  final state in $pp$ collisions at $\sqrt{s}=13$ TeV with the ATLAS
  detector}},
  \href{http://dx.doi.org/10.1016/j.physletb.2018.06.011}{\emph{Phys. Lett. B}
  {\bf 782} (2018) 750--767}, [\href{http://arxiv.org/abs/1803.11145}{{\tt
  1803.11145}}].

\bibitem{ATLAS:2020pcy}
{\scshape ATLAS} collaboration, G.~Aad et~al., \emph{{Search for Higgs Boson
  Decays into a $Z$ Boson and a Light Hadronically Decaying Resonance Using 13
  TeV $pp$ Collision Data from the ATLAS Detector}},
  \href{http://dx.doi.org/10.1103/PhysRevLett.125.221802}{\emph{Phys. Rev.
  Lett.} {\bf 125} (2020) 221802}, [\href{http://arxiv.org/abs/2004.01678}{{\tt
  2004.01678}}].

\bibitem{Cepeda:2021rql}
M.~Cepeda, S.~Gori, V.~M. Outschoorn and J.~Shelton, \emph{{Exotic Higgs
  Decays}},  \href{http://arxiv.org/abs/2111.12751}{{\tt 2111.12751}}.

\bibitem{Englert:2012wf}
C.~Englert, M.~Spannowsky and C.~Wymant, \emph{{Partially (in)visible Higgs
  decays at the LHC}},
  \href{http://dx.doi.org/10.1016/j.physletb.2012.11.008}{\emph{Phys. Lett. B}
  {\bf 718} (2012) 538--544}, [\href{http://arxiv.org/abs/1209.0494}{{\tt
  1209.0494}}].

\bibitem{Petersson:2012dp}
C.~Petersson, A.~Romagnoni and R.~Torre, \emph{{Higgs Decay with Monophoton +
  MET Signature from Low Scale Supersymmetry Breaking}},
  \href{http://dx.doi.org/10.1007/JHEP10(2012)016}{\emph{JHEP} {\bf 10} (2012)
  016}, [\href{http://arxiv.org/abs/1203.4563}{{\tt 1203.4563}}].

\bibitem{CMS:2015ifd}
{\scshape CMS} collaboration, V.~Khachatryan et~al., \emph{{Search for exotic
  decays of a Higgs boson into undetectable particles and one or more
  photons}},
  \href{http://dx.doi.org/10.1016/j.physletb.2015.12.017}{\emph{Phys. Lett. B}
  {\bf 753} (2016) 363--388}, [\href{http://arxiv.org/abs/1507.00359}{{\tt
  1507.00359}}].

\bibitem{CMS:2019ajt}
{\scshape CMS} collaboration, A.~M. Sirunyan et~al., \emph{{Search for dark
  photons in decays of Higgs bosons produced in association with Z bosons in
  proton-proton collisions at $ \sqrt{s} $ = 13 TeV}},
  \href{http://dx.doi.org/10.1007/JHEP10(2019)139}{\emph{JHEP} {\bf 10} (2019)
  139}, [\href{http://arxiv.org/abs/1908.02699}{{\tt 1908.02699}}].

\bibitem{CMS:2020krr}
{\scshape CMS} collaboration, A.~M. Sirunyan et~al., \emph{{Search for dark
  photons in Higgs boson production via vector boson fusion in proton-proton
  collisions at $ \sqrt{s} $ = 13 TeV}},
  \href{http://dx.doi.org/10.1007/JHEP03(2021)011}{\emph{JHEP} {\bf 03} (2021)
  011}, [\href{http://arxiv.org/abs/2009.14009}{{\tt 2009.14009}}].

\bibitem{ATLAS:2021pdg}
{\scshape ATLAS} collaboration, G.~Aad et~al., \emph{{Observation of
  electroweak production of two jets in association with an isolated photon and
  missing transverse momentum, and search for a Higgs boson decaying into
  invisible particles at 13 TeV with the ATLAS detector}},
  \href{http://arxiv.org/abs/2109.00925}{{\tt 2109.00925}}.

\bibitem{ATLAS:2021edm}
{\scshape ATLAS} collaboration, G.~Aad et~al., \emph{{Search for exotic decays
  of the Higgs boson into $b\bar{b}$ and missing transverse momentum in $pp$
  collisions at $\sqrt{s} = 13$ TeV with the ATLAS detector}},
  \href{http://arxiv.org/abs/2109.02447}{{\tt 2109.02447}}.

\bibitem{XENON:2018voc}
{\scshape XENON} collaboration, E.~Aprile et~al., \emph{{Dark Matter Search
  Results from a One Ton-Year Exposure of XENON1T}},
  \href{http://dx.doi.org/10.1103/PhysRevLett.121.111302}{\emph{Phys. Rev.
  Lett.} {\bf 121} (2018) 111302}, [\href{http://arxiv.org/abs/1805.12562}{{\tt
  1805.12562}}].

\bibitem{Boehm:2014hva}
C.~Boehm, M.~J. Dolan, C.~McCabe, M.~Spannowsky and C.~J. Wallace,
  \emph{{Extended gamma-ray emission from Coy Dark Matter}},
  \href{http://dx.doi.org/10.1088/1475-7516/2014/05/009}{\emph{JCAP} {\bf 05}
  (2014) 009}, [\href{http://arxiv.org/abs/1401.6458}{{\tt 1401.6458}}].

\bibitem{Izaguirre:2014vva}
E.~Izaguirre, G.~Krnjaic and B.~Shuve, \emph{{The Galactic Center Excess from
  the Bottom Up}},
  \href{http://dx.doi.org/10.1103/PhysRevD.90.055002}{\emph{Phys. Rev. D} {\bf
  90} (2014) 055002}, [\href{http://arxiv.org/abs/1404.2018}{{\tt 1404.2018}}].

\bibitem{Ipek:2014gua}
S.~Ipek, D.~McKeen and A.~E. Nelson, \emph{{A Renormalizable Model for the
  Galactic Center Gamma Ray Excess from Dark Matter Annihilation}},
  \href{http://dx.doi.org/10.1103/PhysRevD.90.055021}{\emph{Phys. Rev. D} {\bf
  90} (2014) 055021}, [\href{http://arxiv.org/abs/1404.3716}{{\tt 1404.3716}}].

\bibitem{Goodenough:2009gk}
L.~Goodenough and D.~Hooper, \emph{{Possible Evidence For Dark Matter
  Annihilation In The Inner Milky Way From The Fermi Gamma Ray Space
  Telescope}},  \href{http://arxiv.org/abs/0910.2998}{{\tt 0910.2998}}.

\bibitem{Hooper:2010mq}
D.~Hooper and L.~Goodenough, \emph{{Dark Matter Annihilation in The Galactic
  Center As Seen by the Fermi Gamma Ray Space Telescope}},
  \href{http://dx.doi.org/10.1016/j.physletb.2011.02.029}{\emph{Phys. Lett. B}
  {\bf 697} (2011) 412--428}, [\href{http://arxiv.org/abs/1010.2752}{{\tt
  1010.2752}}].

\bibitem{TheFermi-LAT:2015kwa}
{\scshape Fermi-LAT} collaboration, M.~Ajello et~al., \emph{{Fermi-LAT
  Observations of High-Energy $\gamma$-Ray Emission Toward the Galactic
  Center}},
  \href{http://dx.doi.org/10.3847/0004-637X/819/1/44}{\emph{Astrophys. J.} {\bf
  819} (2016) 44}, [\href{http://arxiv.org/abs/1511.02938}{{\tt 1511.02938}}].

\bibitem{Monroe:2007xp}
J.~Monroe and P.~Fisher, \emph{{Neutrino Backgrounds to Dark Matter Searches}},
  \href{http://dx.doi.org/10.1103/PhysRevD.76.033007}{\emph{Phys. Rev. D} {\bf
  76} (2007) 033007}, [\href{http://arxiv.org/abs/0706.3019}{{\tt 0706.3019}}].

\bibitem{deFlorian:2016spz}
{\scshape LHC Higgs Cross Section Working Group} collaboration, D.~de~Florian
  et~al., \emph{{Handbook of LHC Higgs Cross Sections: 4. Deciphering the
  Nature of the Higgs Sector}},  \href{http://arxiv.org/abs/1610.07922}{{\tt
  1610.07922}}.

\bibitem{Alloul:2013bka}
A.~Alloul, N.~D. Christensen, C.~Degrande, C.~Duhr and B.~Fuks,
  \emph{{FeynRules 2.0 - A complete toolbox for tree-level phenomenology}},
  \href{http://dx.doi.org/10.1016/j.cpc.2014.04.012}{\emph{Comput. Phys.
  Commun.} {\bf 185} (2014) 2250--2300},
  [\href{http://arxiv.org/abs/1310.1921}{{\tt 1310.1921}}].

\bibitem{No:2015xqa}
J.~M. No, \emph{{Looking through the pseudoscalar portal into dark matter:
  Novel mono-Higgs and mono-Z signatures at the LHC}},
  \href{http://dx.doi.org/10.1103/PhysRevD.93.031701}{\emph{Phys. Rev. D} {\bf
  93} (2016) 031701}, [\href{http://arxiv.org/abs/1509.01110}{{\tt
  1509.01110}}].

\bibitem{Goncalves:2016iyg}
D.~Goncalves, P.~A.~N. Machado and J.~M. No, \emph{{Simplified Models for Dark
  Matter Face their Consistent Completions}},
  \href{http://dx.doi.org/10.1103/PhysRevD.95.055027}{\emph{Phys. Rev. D} {\bf
  95} (2017) 055027}, [\href{http://arxiv.org/abs/1611.04593}{{\tt
  1611.04593}}].

\bibitem{Bauer:2017ota}
M.~Bauer, U.~Haisch and F.~Kahlhoefer, \emph{{Simplified dark matter models
  with two Higgs doublets: I. Pseudoscalar mediators}},
  \href{http://dx.doi.org/10.1007/JHEP05(2017)138}{\emph{JHEP} {\bf 05} (2017)
  138}, [\href{http://arxiv.org/abs/1701.07427}{{\tt 1701.07427}}].

\bibitem{Alwall:2014hca}
J.~Alwall, R.~Frederix, S.~Frixione, V.~Hirschi, F.~Maltoni, O.~Mattelaer
  et~al., \emph{{The automated computation of tree-level and next-to-leading
  order differential cross sections, and their matching to parton shower
  simulations}}, \href{http://dx.doi.org/10.1007/JHEP07(2014)079}{\emph{JHEP}
  {\bf 07} (2014) 079}, [\href{http://arxiv.org/abs/1405.0301}{{\tt
  1405.0301}}].

\bibitem{Bertone:2017bme}
{\scshape NNPDF} collaboration, V.~Bertone, S.~Carrazza, N.~P. Hartland and
  J.~Rojo, \emph{{Illuminating the photon content of the proton within a global
  PDF analysis}},
  \href{http://dx.doi.org/10.21468/SciPostPhys.5.1.008}{\emph{SciPost Phys.}
  {\bf 5} (2018) 008}, [\href{http://arxiv.org/abs/1712.07053}{{\tt
  1712.07053}}].

\bibitem{Sjostrand:2014zea}
T.~Sj\"ostrand, S.~Ask, J.~R. Christiansen, R.~Corke, N.~Desai, P.~Ilten
  et~al., \emph{{An introduction to PYTHIA 8.2}},
  \href{http://dx.doi.org/10.1016/j.cpc.2015.01.024}{\emph{Comput. Phys.
  Commun.} {\bf 191} (2015) 159--177},
  [\href{http://arxiv.org/abs/1410.3012}{{\tt 1410.3012}}].

\bibitem{deFavereau:2013fsa}
{\scshape DELPHES 3} collaboration, J.~de~Favereau, C.~Delaere, P.~Demin,
  A.~Giammanco, V.~Lema\^\i{}tre, A.~Mertens et~al., \emph{{DELPHES 3, A
  modular framework for fast simulation of a generic collider experiment}},
  \href{http://dx.doi.org/10.1007/JHEP02(2014)057}{\emph{JHEP} {\bf 02} (2014)
  057}, [\href{http://arxiv.org/abs/1307.6346}{{\tt 1307.6346}}].

\bibitem{2008:triboson}
T.~Binoth, G.~Ossola, C.~Papadopoulos and R.~Pittau, \emph{Nlo qcd corrections
  to tri-boson production},
  \href{http://dx.doi.org/10.1088/1126-6708/2008/06/082}{\emph{Journal of High
  Energy Physics} {\bf 2008} (Jun, 2008) 082–082}.

\bibitem{2019:ttz}
A.~Kulesza, L.~Motyka, D.~Schwartländer, T.~Stebel and V.~Theeuwes,
  \emph{Associated production of a top quark pair with a heavy electroweak
  gauge boson at nlo+nnll accuracy},
  \href{http://dx.doi.org/10.1140/epjc/s10052-019-6746-z}{\emph{The European
  Physical Journal C} {\bf 79} (Mar, 2019) }.

\bibitem{2014:zz}
F.~Cascioli, T.~Gehrmann, M.~Grazzini, S.~Kallweit, P.~Maierhöfer, A.~von
  Manteuffel et~al., \emph{Zz production at hadron colliders in nnlo qcd},
  \href{http://dx.doi.org/10.1016/j.physletb.2014.06.056}{\emph{Physics Letters
  B} {\bf 735} (Jul, 2014) 311–313}.

\bibitem{Frixione:2008yi}
S.~Frixione, E.~Laenen, P.~Motylinski, B.~R. Webber and C.~D. White,
  \emph{{Single-top hadroproduction in association with a W boson}},
  \href{http://dx.doi.org/10.1088/1126-6708/2008/07/029}{\emph{JHEP} {\bf 07}
  (2008) 029}, [\href{http://arxiv.org/abs/0805.3067}{{\tt 0805.3067}}].

\bibitem{Faham:2021zet}
H.~E. Faham, F.~Maltoni, K.~Mimasu and M.~Zaro, \emph{{Single top production in
  association with a WZ pair at the LHC in the SMEFT}},
  \href{http://dx.doi.org/10.1007/JHEP01(2022)100}{\emph{JHEP} {\bf 01} (2022)
  100}, [\href{http://arxiv.org/abs/2111.03080}{{\tt 2111.03080}}].

\bibitem{ATL-DAQ-PUB-2019-001}
{\scshape ATLAS} collaboration, \emph{{Trigger menu in 2018}},
  {\emph{ATL-DAQ-PUB-2019-001} (Oct, 2019) }.

\bibitem{2011Cowan}
G.~Cowan, K.~Cranmer, E.~Gross and O.~Vitells, \emph{Asymptotic formulae for
  likelihood-based tests of new physics},
  \href{http://dx.doi.org/10.1140/epjc/s10052-011-1554-0}{\emph{The European
  Physical Journal C} {\bf 71} (Feb, 2011) }.

\bibitem{Behnke:2013xla}
\emph{{The International Linear Collider Technical Design Report - Volume 1:
  Executive Summary}},  \href{http://arxiv.org/abs/1306.6327}{{\tt 1306.6327}}.

\bibitem{bambade2019international}
P.~Bambade, T.~Barklow, T.~Behnke, M.~Berggren, J.~Brau, P.~Burrows et~al.,
  \emph{The international linear collider: A global project},
  \href{http://arxiv.org/abs/1903.01629}{{\tt 1903.01629}}.

\bibitem{Potter:2016pgp}
C.~T. Potter, \emph{{DSiD: a Delphes Detector for ILC Physics Studies}},  in
  \emph{{International Workshop on Future Linear Colliders}}, 2, 2016.
\newblock \href{http://arxiv.org/abs/1602.07748}{{\tt 1602.07748}}.

\bibitem{Brivio:2017ije}
I.~Brivio, M.~B. Gavela, L.~Merlo, K.~Mimasu, J.~M. No, R.~del Rey et~al.,
  \emph{{ALPs Effective Field Theory and Collider Signatures}},
  \href{http://dx.doi.org/10.1140/epjc/s10052-017-5111-3}{\emph{Eur. Phys. J.
  C} {\bf 77} (2017) 572}, [\href{http://arxiv.org/abs/1701.05379}{{\tt
  1701.05379}}].

\bibitem{Bauer:2017ris}
M.~Bauer, M.~Neubert and A.~Thamm, \emph{{Collider Probes of Axion-Like
  Particles}}, \href{http://dx.doi.org/10.1007/JHEP12(2017)044}{\emph{JHEP}
  {\bf 12} (2017) 044}, [\href{http://arxiv.org/abs/1708.00443}{{\tt
  1708.00443}}].

\bibitem{Dolan:2017osp}
M.~J. Dolan, T.~Ferber, C.~Hearty, F.~Kahlhoefer and K.~Schmidt-Hoberg,
  \emph{{Revised constraints and Belle II sensitivity for visible and invisible
  axion-like particles}},
  \href{http://dx.doi.org/10.1007/JHEP12(2017)094}{\emph{JHEP} {\bf 12} (2017)
  094}, [\href{http://arxiv.org/abs/1709.00009}{{\tt 1709.00009}}].

\bibitem{Alves:2019xpc}
A.~Alves, A.~G. Dias and D.~D. Lopes, \emph{{Probing ALP-Sterile Neutrino
  Couplings at the LHC}},
  \href{http://dx.doi.org/10.1007/JHEP08(2020)074}{\emph{JHEP} {\bf 08} (2020)
  074}, [\href{http://arxiv.org/abs/1911.12394}{{\tt 1911.12394}}].

\bibitem{Alonso-Alvarez:2018irt}
G.~Alonso-\'Alvarez, M.~B. Gavela and P.~Quilez, \emph{{Axion couplings to
  electroweak gauge bosons}},
  \href{http://dx.doi.org/10.1140/epjc/s10052-019-6732-5}{\emph{Eur. Phys. J.
  C} {\bf 79} (2019) 223}, [\href{http://arxiv.org/abs/1811.05466}{{\tt
  1811.05466}}].

\bibitem{Planck:2015fie}
{\scshape Planck} collaboration, P.~A.~R. Ade et~al., \emph{{Planck 2015
  results. XIII. Cosmological parameters}},
  \href{http://dx.doi.org/10.1051/0004-6361/201525830}{\emph{Astron.
  Astrophys.} {\bf 594} (2016) A13},
  [\href{http://arxiv.org/abs/1502.01589}{{\tt 1502.01589}}].

\bibitem{Robens:2021lov}
T.~Robens, \emph{{The THDMa Revisited}},
  \href{http://dx.doi.org/10.3390/sym13122341}{\emph{Symmetry} {\bf 13} (2021)
  2341}, [\href{http://arxiv.org/abs/2106.02962}{{\tt 2106.02962}}].

\bibitem{Ertas:2019dew}
F.~Ertas and F.~Kahlhoefer, \emph{{Loop-induced direct detection signatures
  from CP-violating scalar mediators}},
  \href{http://dx.doi.org/10.1007/JHEP06(2019)052}{\emph{JHEP} {\bf 06} (2019)
  052}, [\href{http://arxiv.org/abs/1902.11070}{{\tt 1902.11070}}].

\bibitem{Abe:2019wjw}
T.~Abe, M.~Fujiwara, J.~Hisano and Y.~Shoji, \emph{{Maximum value of the
  spin-independent cross section in the 2HDM+a}},
  \href{http://dx.doi.org/10.1007/JHEP01(2020)114}{\emph{JHEP} {\bf 01} (2020)
  114}, [\href{http://arxiv.org/abs/1910.09771}{{\tt 1910.09771}}].

\bibitem{Tunney:2017yfp}
P.~Tunney, J.~M. No and M.~Fairbairn, \emph{{Probing the pseudoscalar portal to
  dark matter via $\bar bbZ(\to\ell\ell)+ \not{E}_T$ : From the LHC to the
  Galactic Center excess}},
  \href{http://dx.doi.org/10.1103/PhysRevD.96.095020}{\emph{Phys. Rev. D} {\bf
  96} (2017) 095020}, [\href{http://arxiv.org/abs/1705.09670}{{\tt
  1705.09670}}].

\bibitem{LHCDarkMatterWorkingGroup:2018ufk}
{\scshape LHC Dark Matter Working Group} collaboration, T.~Abe et~al.,
  \emph{{LHC Dark Matter Working Group: Next-generation spin-0 dark matter
  models}}, \href{http://dx.doi.org/10.1016/j.dark.2019.100351}{\emph{Phys.
  Dark Univ.} {\bf 27} (2020) 100351},
  [\href{http://arxiv.org/abs/1810.09420}{{\tt 1810.09420}}].

\bibitem{Gunion:2002zf}
J.~F. Gunion and H.~E. Haber, \emph{{The CP conserving two Higgs doublet model:
  The Approach to the decoupling limit}},
  \href{http://dx.doi.org/10.1103/PhysRevD.67.075019}{\emph{Phys. Rev. D} {\bf
  67} (2003) 075019}, [\href{http://arxiv.org/abs/hep-ph/0207010}{{\tt
  hep-ph/0207010}}].

\bibitem{Gunion:1989we}
J.~F. Gunion, H.~E. Haber, G.~L. Kane and S.~Dawson, \emph{{The Higgs Hunter's
  Guide}}, vol.~80.
\newblock 2000.

\bibitem{Branco:2011iw}
G.~Branco, P.~Ferreira, L.~Lavoura, M.~Rebelo, M.~Sher and J.~P. Silva,
  \emph{{Theory and phenomenology of two-Higgs-doublet models}},
  \href{http://dx.doi.org/10.1016/j.physrep.2012.02.002}{\emph{Phys. Rept.}
  {\bf 516} (2012) 1--102}, [\href{http://arxiv.org/abs/1106.0034}{{\tt
  1106.0034}}].

\bibitem{CMS:2020gsy}
{\scshape CMS} collaboration, \emph{{Combined Higgs boson production and decay
  measurements with up to 137 fb-1 of proton-proton collision data at sqrts =
  13 TeV}}, {\emph{CMS-PAS-HIG-19-005} (1, 2020) }.

\bibitem{ATLAS:2020qdt}
{\scshape ATLAS} collaboration, \emph{{A combination of measurements of Higgs
  boson production and decay using up to $139$ fb$^{-1}$ of proton--proton
  collision data at $\sqrt{s}=$ 13 TeV collected with the ATLAS experiment}},
  {\emph{ATLAS-CONF-2020-027} (8, 2020) }.

\bibitem{TheATLAScollaboration:2014qxe}
{\scshape ATLAS} collaboration, \emph{{Prospects for New Physics in Higgs
  Couplings Studies with the ATLAS Detector at the HL-LHC}},
  {\emph{ATL-PHYS-PUB-2014-017} (10, 2014) }.

\bibitem{Muhlleitner:2020wwk}
M.~M\"uhlleitner, M.~O.~P. Sampaio, R.~Santos and J.~Wittbrodt,
  \emph{{ScannerS: parameter scans in extended scalar sectors}},
  \href{http://dx.doi.org/10.1140/epjc/s10052-022-10139-w}{\emph{Eur. Phys. J.
  C} {\bf 82} (2022) 198}, [\href{http://arxiv.org/abs/2007.02985}{{\tt
  2007.02985}}].

\bibitem{Bechtle:2013xfa}
P.~Bechtle, S.~Heinemeyer, O.~St\r{a}l, T.~Stefaniak and G.~Weiglein,
  \emph{{$HiggsSignals$: Confronting arbitrary Higgs sectors with measurements
  at the Tevatron and the LHC}},
  \href{http://dx.doi.org/10.1140/epjc/s10052-013-2711-4}{\emph{Eur. Phys. J.
  C} {\bf 74} (2014) 2711}, [\href{http://arxiv.org/abs/1305.1933}{{\tt
  1305.1933}}].

\bibitem{Bechtle:2020uwn}
P.~Bechtle, S.~Heinemeyer, T.~Klingl, T.~Stefaniak, G.~Weiglein and
  J.~Wittbrodt, \emph{{HiggsSignals-2: Probing new physics with precision Higgs
  measurements in the LHC 13 TeV era}},
  \href{http://dx.doi.org/10.1140/epjc/s10052-021-08942-y}{\emph{Eur. Phys. J.
  C} {\bf 81} (2021) 145}, [\href{http://arxiv.org/abs/2012.09197}{{\tt
  2012.09197}}].

\bibitem{Fabbrichesi:2020wbt}
M.~Fabbrichesi, E.~Gabrielli and G.~Lanfranchi, \emph{{The Dark Photon}},
  \href{http://arxiv.org/abs/2005.01515}{{\tt 2005.01515}}.

\bibitem{Jaeckel:2012yz}
J.~Jaeckel, M.~Jankowiak and M.~Spannowsky, \emph{{LHC probes the hidden
  sector}}, \href{http://dx.doi.org/10.1016/j.dark.2013.06.001}{\emph{Phys.
  Dark Univ.} {\bf 2} (2013) 111--117},
  [\href{http://arxiv.org/abs/1212.3620}{{\tt 1212.3620}}].

\bibitem{Hook:2010tw}
A.~Hook, E.~Izaguirre and J.~G. Wacker, \emph{{Model Independent Bounds on
  Kinetic Mixing}}, \href{http://dx.doi.org/10.1155/2011/859762}{\emph{Adv.
  High Energy Phys.} {\bf 2011} (2011) 859762},
  [\href{http://arxiv.org/abs/1006.0973}{{\tt 1006.0973}}].

\bibitem{Fox:2011fx}
P.~J. Fox, R.~Harnik, J.~Kopp and Y.~Tsai, \emph{{LEP Shines Light on Dark
  Matter}}, \href{http://dx.doi.org/10.1103/PhysRevD.84.014028}{\emph{Phys.
  Rev. D} {\bf 84} (2011) 014028}, [\href{http://arxiv.org/abs/1103.0240}{{\tt
  1103.0240}}].

\bibitem{DELPHI:2008uka}
{\scshape DELPHI} collaboration, J.~Abdallah et~al., \emph{{Search for one
  large extra dimension with the DELPHI detector at LEP}},
  \href{http://dx.doi.org/10.1140/epjc/s10052-009-0874-9}{\emph{Eur. Phys. J.
  C} {\bf 60} (2009) 17--23}, [\href{http://arxiv.org/abs/0901.4486}{{\tt
  0901.4486}}].

\bibitem{BaBar:2010eww}
{\scshape BaBar} collaboration, P.~del Amo~Sanchez et~al., \emph{{Search for
  Production of Invisible Final States in Single-Photon Decays of
  $\Upsilon(1S)$}},
  \href{http://dx.doi.org/10.1103/PhysRevLett.107.021804}{\emph{Phys. Rev.
  Lett.} {\bf 107} (2011) 021804}, [\href{http://arxiv.org/abs/1007.4646}{{\tt
  1007.4646}}].

\bibitem{BaBar:2008aby}
{\scshape BaBar} collaboration, B.~Aubert et~al., \emph{{Search for Invisible
  Decays of a Light Scalar in Radiative Transitions $\upsilon_{3S} \to \gamma$
  A0}},  in \emph{{34th International Conference on High Energy Physics}}, 7,
  2008.
\newblock \href{http://arxiv.org/abs/0808.0017}{{\tt 0808.0017}}.

\bibitem{CrystalBall:1990xec}
{\scshape Crystal Ball} collaboration, D.~Antreasyan et~al., \emph{{Limits on
  axion and light Higgs boson production in Upsilon (1s) decays}},
  \href{http://dx.doi.org/10.1016/0370-2693(90)90254-4}{\emph{Phys. Lett. B}
  {\bf 251} (1990) 204--210}.

\bibitem{Knapen:2016moh}
S.~Knapen, T.~Lin, H.~K. Lou and T.~Melia, \emph{{Searching for Axionlike
  Particles with Ultraperipheral Heavy-Ion Collisions}},
  \href{http://dx.doi.org/10.1103/PhysRevLett.118.171801}{\emph{Phys. Rev.
  Lett.} {\bf 118} (2017) 171801}, [\href{http://arxiv.org/abs/1607.06083}{{\tt
  1607.06083}}].

\bibitem{ATLAS:2020hii}
{\scshape ATLAS} collaboration, G.~Aad et~al., \emph{{Measurement of
  light-by-light scattering and search for axion-like particles with 2.2
  nb$^{-1}$ of Pb+Pb data with the ATLAS detector}},
  \href{http://dx.doi.org/10.1007/JHEP11(2021)050}{\emph{JHEP} {\bf 11} (2021)
  050}, [\href{http://arxiv.org/abs/2008.05355}{{\tt 2008.05355}}].

\bibitem{CMS:2018erd}
{\scshape CMS} collaboration, A.~M. Sirunyan et~al., \emph{{Evidence for
  light-by-light scattering and searches for axion-like particles in
  ultraperipheral PbPb collisions at $\sqrt{s_\mathrm{NN}} =$ 5.02 TeV}},
  \href{http://dx.doi.org/10.1016/j.physletb.2019.134826}{\emph{Phys. Lett. B}
  {\bf 797} (2019) 134826}, [\href{http://arxiv.org/abs/1810.04602}{{\tt
  1810.04602}}].

\bibitem{Bonilla:2022pxu}
J.~Bonilla, I.~Brivio, J.~Machado-Rodr\'\i{}guez and J.~F. de~Troc\'oniz,
  \emph{{Nonresonant Searches for Axion-Like Particles in Vector Boson
  Scattering Processes at the LHC}},
  \href{http://arxiv.org/abs/2202.03450}{{\tt 2202.03450}}.

\bibitem{Raffelt:2006cw}
G.~G. Raffelt, \emph{{Astrophysical axion bounds}},
  \href{http://dx.doi.org/10.1007/978-3-540-73518-2_3}{\emph{Lect. Notes Phys.}
  {\bf 741} (2008) 51--71}, [\href{http://arxiv.org/abs/hep-ph/0611350}{{\tt
  hep-ph/0611350}}].

\bibitem{Mimasu:2014nea}
K.~Mimasu and V.~Sanz, \emph{{ALPs at Colliders}},
  \href{http://dx.doi.org/10.1007/JHEP06(2015)173}{\emph{JHEP} {\bf 06} (2015)
  173}, [\href{http://arxiv.org/abs/1409.4792}{{\tt 1409.4792}}].

\bibitem{Jaeckel:2015jla}
J.~Jaeckel and M.~Spannowsky, \emph{{Probing MeV to 90 GeV axion-like particles
  with LEP and LHC}},
  \href{http://dx.doi.org/10.1016/j.physletb.2015.12.037}{\emph{Phys. Lett. B}
  {\bf 753} (2016) 482--487}, [\href{http://arxiv.org/abs/1509.00476}{{\tt
  1509.00476}}].

\bibitem{Dobrich:2015jyk}
B.~D\"obrich, J.~Jaeckel, F.~Kahlhoefer, A.~Ringwald and K.~Schmidt-Hoberg,
  \emph{{ALPtraum: ALP production in proton beam dump experiments}},
  \href{http://dx.doi.org/10.1007/JHEP02(2016)018}{\emph{JHEP} {\bf 02} (2016)
  018}, [\href{http://arxiv.org/abs/1512.03069}{{\tt 1512.03069}}].

\bibitem{L3:1997exg}
{\scshape L3} collaboration, M.~Acciarri et~al., \emph{{Search for new physics
  in energetic single photon production in $e^{+} e^{-}$ annihilation at the
  $Z$ resonance}},
  \href{http://dx.doi.org/10.1016/S0370-2693(97)01003-4}{\emph{Phys. Lett. B}
  {\bf 412} (1997) 201--209}.

\bibitem{Brooijmans:2020yij}
G.~Brooijmans et~al., \emph{{Les Houches 2019 Physics at TeV Colliders: New
  Physics Working Group Report}},  in \emph{{11th Les Houches Workshop on
  Physics at TeV Colliders}: {PhysTeV Les Houches}}, 2, 2020.
\newblock \href{http://arxiv.org/abs/2002.12220}{{\tt 2002.12220}}.

\bibitem{ATLAS:2020kdi}
{\scshape ATLAS} collaboration, \emph{{Combination of searches for invisible
  Higgs boson decays with the ATLAS experiment}}, {\emph{ATLAS-CONF-2020-052}
  (10, 2020) }.

\bibitem{CMS:2018tip}
{\scshape CMS} collaboration, \emph{{Search for invisible decays of a Higgs
  boson produced through vector boson fusion at the High-Luminosity LHC}},
  {\emph{CMS-PAS-FTR-18-016} (2018) }.

\bibitem{CMS:2021snz}
{\scshape CMS} collaboration, \emph{{Search for new particles in events with
  energetic jets and large missing transverse momentum in proton-proton
  collisions at $\sqrt{s}=13~\mathrm{TeV}$}}, {\emph{CMS-PAS-EXO-20-004} (2021)
  }.

\bibitem{Abbiendi:2002yk}
{\scshape OPAL} collaboration, G.~Abbiendi et~al., \emph{{Search for the
  standard model Higgs boson with the OPAL detector at LEP}},
  \href{http://dx.doi.org/10.1140/epjc/s2002-01092-3}{\emph{Eur. Phys. J. C}
  {\bf 26} (2003) 479--503}, [\href{http://arxiv.org/abs/hep-ex/0209078}{{\tt
  hep-ex/0209078}}].

\bibitem{Achard:2001pj}
{\scshape L3} collaboration, P.~Achard et~al., \emph{{Standard model Higgs
  boson with the L3 experiment at LEP}},
  \href{http://dx.doi.org/10.1016/S0370-2693(01)01010-3}{\emph{Phys. Lett. B}
  {\bf 517} (2001) 319--331}, [\href{http://arxiv.org/abs/hep-ex/0107054}{{\tt
  hep-ex/0107054}}].

\bibitem{Heister:2001kr}
{\scshape ALEPH} collaboration, A.~Heister et~al., \emph{{Final results of the
  searches for neutral Higgs bosons in e+ e- collisions at s**(1/2) up to
  209-GeV}}, \href{http://dx.doi.org/10.1016/S0370-2693(01)01487-3}{\emph{Phys.
  Lett. B} {\bf 526} (2002) 191--205},
  [\href{http://arxiv.org/abs/hep-ex/0201014}{{\tt hep-ex/0201014}}].

\bibitem{Barate:2000ts}
{\scshape ALEPH} collaboration, R.~Barate et~al., \emph{{Observation of an
  excess in the search for the standard model Higgs boson at ALEPH}},
  \href{http://dx.doi.org/10.1016/S0370-2693(00)01269-7}{\emph{Phys. Lett. B}
  {\bf 495} (2000) 1--17}, [\href{http://arxiv.org/abs/hep-ex/0011045}{{\tt
  hep-ex/0011045}}].

\bibitem{Abdallah:2003ip}
{\scshape DELPHI} collaboration, J.~Abdallah et~al., \emph{{Final results from
  DELPHI on the searches for SM and MSSM neutral Higgs bosons}},
  \href{http://dx.doi.org/10.1140/epjc/s2003-01394-x}{\emph{Eur. Phys. J. C}
  {\bf 32} (2004) 145--183}, [\href{http://arxiv.org/abs/hep-ex/0303013}{{\tt
  hep-ex/0303013}}].

\bibitem{Hermann:2012fc}
T.~Hermann, M.~Misiak and M.~Steinhauser, \emph{{$\bar{B}\to X_s \gamma$ in the
  Two Higgs Doublet Model up to Next-to-Next-to-Leading Order in QCD}},
  \href{http://dx.doi.org/10.1007/JHEP11(2012)036}{\emph{JHEP} {\bf 11} (2012)
  036}, [\href{http://arxiv.org/abs/1208.2788}{{\tt 1208.2788}}].

\bibitem{Misiak:2020vlo}
M.~Misiak, A.~Rehman and M.~Steinhauser, \emph{{Towards $ \overline{B}\to
  {X}_s\gamma $ at the NNLO in QCD without interpolation in m$_{c}$}},
  \href{http://dx.doi.org/10.1007/JHEP06(2020)175}{\emph{JHEP} {\bf 06} (2020)
  175}, [\href{http://arxiv.org/abs/2002.01548}{{\tt 2002.01548}}].

\bibitem{Skiba:1992mg}
W.~Skiba and J.~Kalinowski, \emph{{$B_s \to \tau^{+} \tau^{-}$ decay in a two
  Higgs doublet model}},
  \href{http://dx.doi.org/10.1016/0550-3213(93)90470-A}{\emph{Nucl. Phys. B}
  {\bf 404} (1993) 3--19}.

\bibitem{Logan:2000iv}
H.~E. Logan and U.~Nierste, \emph{{$B_{s,d} \to \ell^+ \ell^-$ in a two Higgs
  doublet model}},
  \href{http://dx.doi.org/10.1016/S0550-3213(00)00417-X}{\emph{Nucl. Phys. B}
  {\bf 586} (2000) 39--55}, [\href{http://arxiv.org/abs/hep-ph/0004139}{{\tt
  hep-ph/0004139}}].

\bibitem{ATLAS:2021gcn}
{\scshape ATLAS} collaboration, G.~Aad et~al., \emph{{Search for associated
  production of a Z boson with an invisibly decaying Higgs boson or dark matter
  candidates at s=13 TeV with the ATLAS detector}},
  \href{http://dx.doi.org/10.1016/j.physletb.2022.137066}{\emph{Phys. Lett. B}
  {\bf 829} (2022) 137066}, [\href{http://arxiv.org/abs/2111.08372}{{\tt
  2111.08372}}].

\bibitem{ATLAS:2020tlo}
{\scshape ATLAS} collaboration, G.~Aad et~al., \emph{{Search for heavy
  resonances decaying into a pair of Z bosons in the $\ell ^+\ell ^-\ell
  '^+\ell '^-$ and $\ell ^+\ell ^-\nu {{\bar{\nu }}}$ final states using 139
  $\mathrm {fb}^{-1}$ of proton\textendash{}proton collisions at $\sqrt{s} =
  13\,$TeV with the ATLAS detector}},
  \href{http://dx.doi.org/10.1140/epjc/s10052-021-09013-y}{\emph{Eur. Phys. J.
  C} {\bf 81} (2021) 332}, [\href{http://arxiv.org/abs/2009.14791}{{\tt
  2009.14791}}].

\end{thebibliography}\endgroup

\end{document}